\documentclass[twocol]{ametsocV6.1}

\def\bm#1{\boldsymbol{#1}}

\def\av#1{\langle#1\rangle}

\def\mbf#1{\mbox{\textbf{#1}}}

\title{On idealized models of turbulent condensation in clouds}

\authors{Gustavo C. Abade\correspondingauthor{Gustavo Abade, gustavo.abade@fuw.edu.pl}}

\affiliation{Institute of Geophysics, Faculty of Physics, University of Warsaw, Poland}

\abstract{Various microphysical models attempt to explain the
  occurrence of broad droplet size distributions (DSD) in clouds
  through approximate representations of the stochastic droplet growth
  by condensation in a turbulent environment. This work analyzes
  specific idealized models, where the variability of droplet growth
  conditions arises primarily from variability in the turbulent
  vertical velocity of the air carrying these droplets. Examples are
  the stochastic eddy hopping model operating in adiabatic parcels and
  certain types of DNS-like models. We show that such models produce
  droplet size statistics that are spatially inhomogeneous along the
  vertical direction, causing the predicted DSD to depend on the DSD
  spatial sampling scale $\Delta$. In these models, $\Delta$ is
  implicitly related to the spatial extent of the droplets turbulent
  diffusion (approximated by Brownian-like excursions) and thus grows
  like $t^{1/2}$. This leads to spurious continuous DSD broadening, as
  the growth in time (also like $t^{1/2}$) of the standard deviation
  of droplet squared radius arises essentially from the growth of the
  sampling scale $\Delta$. Also, the DSDs predicted by the models
  discussed here are not necessarily \textit{locally} broad (in the
  sense that large and small droplets might not be well-mixed in
  sufficiently small volumes) and thus do not necessarily indicate 
  enhanced probabilities of gravity-induced droplet coagulation. In
  the effort to build a firm physical basis for subgrid
  parametrizations, this study presents a framework to explain the
  merits and limitations of idealized models, indicating how to assess
  and use them wisely as a subgrid representation of turbulent
  condensation in large-eddy simulations of clouds.}

\begin{document}

\maketitle

%
%
%
\statement

The droplet size distribution is a prime characteristic of a cloud, controlling the initiation of precipitation and how the cloud modulates solar radiation. Microphysical processes shaping the droplet size distribution, such as droplet growth by condensation in the turbulent cloud flow, occur on scales that are small compared with the resolution of the numerical models used to simulate the entire cloud. Such subgrid-scale aspects are incorporated in these models by parametrizations relating microphysical processes to the model variables. In the effort to build a firm physical basis for such parametrizations, this study analyzes idealized representations of turbulent condensation at subgrid-scales and presents a framework to explain the merits and limitations of these representations, indicating how to overcome their deficiencies.

\section{Introduction}

Adiabatic cloud parcel models and direct numerical simulations (DNS)
of cloudy volumes with triply periodic boundary conditions (PBC) are
widely used frameworks to study cloud microphysics. These idealized
frameworks are designed to provide basic understanding of cloud
processes occurring at small scales and eventually give insights into
the central problem of subgrid-scale (SGS) modeling of cloud
microphysics in large-eddy simulations (LES) of clouds.

{\color{black}{The suitability of these idealized models to explain,
    in particular, the occurrence of local broad droplet size
    distributions in clouds has been recently questioned
    by~\cite{Prabhakaran_et_al_2022}}}. Their objections concern
models where the variability of droplet growth conditions (determined
by the supersaturation droplets experience) is due primarily to
variability of the turbulent vertical velocity of the air carrying
these droplets (through the adiabatic cooling effect). Examples of
such models are the so-called eddy hopping model
[e.g.,~\citet{Sardina_et_al_2015,Grabowski_Abade_2017,Abade_et_al_2018}]
operating in adiabatic cloud parcels and a certain type of DNS-like
models
[e.g.,~\citet{Vaillancourt_et_al_2002,Lanotte_et_al_2009,Sardina_et_al_2015,Sardina_et_al_2018,Saito_and_Gotoh_2018,Thomas_et_al_2020,Grabowski_et_al_DNS_CCN_single,Grabowski_et_al_DNS_CCN_distr}].

The models mentioned above {\color{black}{presumably}} predict broad droplet size distributions,
in general qualitative agreement with \textit{in situ}
measurements. However, as we shall explain later, the predicted
droplet size distributions are dependent on the sampling volume (which
is generally unspecified) and cannot be generally regarded as \textit{locally}
broad. That is, large and small droplets, forming the left and right
tails of the predicted distribution, {\color{black}{might not be}} well-mixed in
sufficiently small cloudy volumes. This feature is important because
locally broad droplet size distributions accelerate the onset of the
collision-coalescence process, as they enhance the probability of
gravity-induced droplet collision due to differential sedimentation
velocities of droplets having different sizes.

Idealized adiabatic cloud parcels and cloudy periodic boxes in DNS
models, combined with some explicit treatment of the microphysics, are
commonly regarded as instrumental models to represent the unresolved
subgrid cloudy states in LES. This is based on the picture that the
subgrid microphysical state of a particular grid box (of linear size
$\Delta$) consists of a statistical ensemble of the microphysical
states of adiabatic parcels or DNS periodic boxes (of characteristic
size $\Delta$) that instantaneously occupy the volume of this
particular grid box. In this picture, there are two essential aspects
associated with the characteristic size $\Delta$: (a) this is
generally the scale on which the cloudy air in a grid box is assumed
to acquire its mean turbulent kinetic energy $k$ at the rate
$\varepsilon$ per unit mass; and (b) it specifies the linear extent of
the \textit{sampling volume} over which relevant microphysical
characteristics, such as the droplet size distribution and its
moments, are evaluated. One way to state the central issue we bring
out in this paper is that the eddy hopping model and some DNS-like
studies observe only the first aspect (a), that is, $\Delta$ as the
scale on which turbulent kinetic energy is supplied to the system, but
inadvertently miss the second essential aspect (b), that $\Delta$ also
specifies the extent of the averaging volume. This omission results in
``non-local" sampling of microphysical properties and arises from
ill-suited boundary conditions, as well shall clarify later in this
work. {\color{black}{Issues with ill-suited boundary conditions used in these idealized models were indicated by~\citet{Prabhakaran_et_al_2022}.}} 

This work describes a framework to explain the deficiencies and
merits of the idealized models of turbulent condensation mentioned
above. The suggested framework addresses the specific comments
of~\cite{Prabhakaran_et_al_2022} (see their Section~5.b) and assists
us in properly sampling the droplet size statistics in idealized models, as we shall do for
the eddy hopping model operating in adiabatic cloud parcels. In
particular, understanding the deficiencies and merits of the eddy
hopping model (as a stand-alone representation of turbulent
condensation in adiabatic parcels) will motivate its wise use as a
subgrid condensation model built in an LES scheme that employs
particle-based microphysics. We also formulate a consistency condition
for any subgrid model of turbulent condensation that (explicitly or
not) includes turbulent vertical velocity fluctuations as one possible
source of subgrid supersaturation variability.

Before we proceed with this program, some preliminary background is
provided in Sec.~\ref{sec:dsd}, where we define the droplet size
distribution (DSD) and discuss its dependence on the sampling
volume. For clarity we outline in Sec.~\ref{sec:cooper} the
theoretical framework of~\citet{Cooper_1989}, which is based on the
quasi-steady supersaturation approximation, and recall the
\textit{reversible condensation} scenario. This is a useful reference
scenario where no strictly local DSD broadening is possible. The
framework of~\citet{Cooper_1989} incorporates many previous
explanations~[e.g.,~\citet{bartlett_jonas_qj_1972,Manton_qjrms_1979}]
of the flaws in earlier versions of the stochastic condensation
theory, which was revised and improved by~\citet{KC_1999a}.  In
Section~\ref{sec:parcel} we recall the essentials of the eddy hopping
model operating in adiabatic parcels, explain its deficiencies and
propose {\color{black}{ways to enforce a correct \textit{local sampling} of the DSD}}. Section~\ref{sec:clark_hall}
discusses deviations from the quasi-steady approximation and explores
insights from the work of~\citet{Clark_and_Hall_1979} on possible ways
in which DSD may become locally broad. Section~\ref{sec:dns} briefly reviews two types of
DNS-like studies of turbulent condensation and explain that one of
these types has the same deficiencies of the eddy hopping model. In Section~\ref{sec:sgs_model}
we discuss some aspects that emerge when idealized stochastic models
(in particular, the eddy hopping model) are used as a subgrid
condensation scheme in LES of clouds with particle-based
microphysics. In
Sec.~\ref{sec:conclusions} we summarize the main points of this work.

\section{Fine-grained and filtered DSD}
\label{sec:dsd}

The fundamental characteristic of a cloudy volume considered in this
work is the droplet size distribution (DSD). Because it may depend on
the volume over which the distribution is averaged, every well-defined
DSD goes with the specification of its underlying spatial averaging
scale.

Here we define the \textit{local} or fine-grained DSD $n(r;\mbf x,t)$
in such a way that
\begin{equation}
n(r; \mbf x, t) \: d r d \mbf x,
\label{eq:dsd_def}
\end{equation}
is the number of droplets that lie in the volume element $d \mbf x$
around $\mbf x$ at time $t$ and that have a radius in the range $[r, r
  + dr]$. This definition implicitly incorporates the
\textit{continuum hypothesis} for the disperse liquid phase in the
cloudy air. Thus \textit{local} refers to the mathematically
infinitesimal volume element $d \mbf x$ of a particular ``physical
point" in the continuous cloud. This physical point is assumed to be
``macroscopically infinitesimal", but at the same time
``mesoscopically infinite" [following analogous considerations
  by~\citet{Resibois_book} while defining regular distribution
  functions used in the kinetic theory of gases]. This means that such
a physical point is small enough for all microphysical properties of
the cloudy air not vary appreciably over the extension of the sampling
volume $d \mbf x$, but at the same time big enough to contain a large
number of droplets.

The fine-grained DSD discussed above is thus the local size
distribution in a continuous description of cloudy air. However, the
continuous equations of cloudy air motion are intractable except by
numerical methods. These methods require spatial discretization that
naturally introduces a cutoff lenghtscale $\Delta$ determining the
finite spatial resolution of the numerical model. In an LES setting,
$\Delta$ may be loosely identified with the size of the computational
grid box. A grid box is considered a ``point" in the spatially
discretized LES domain. Thus the ``local" DSD at this ``grid point" is
actually a spatially filtered DSD,
\begin{equation}
\av{n(r;\mbf x,t)}_{\Delta} = \int G(\mbf x - \mbf x') n(r;\mbf x',t) d \mbf x',
\label{eq:filtered_dsd_def}
\end{equation}
where $G$ is a spatial filter function of characteristic filter width
$\Delta$ and satisfies $\int G(\mbf x) d \mbf x = 1$ [see e.g.,~\citet{Germano_1992}]. (In the limit of
$\Delta \to 0$, the filter function $G$ tends to the Dirac delta
distribution.) We note that LES generally does not define the filter
function $G$ explicitly; an implicit filter arises from the spatial
discretization. Thus the distribution $\av{n(r)}_{\Delta}$ is simply
the grid average (or the $G$-weighted spatial average) of the
fine-grained $n(r)$, where $\Delta$ is the spatial averaging
lengthscale. Also, we define $\mu_{k} = \int r^{k}n(r)dr$ as the
$k$th moment of the fine-grained DSD $n(r)$ and
$\av{\mu_{k}}_{\Delta} = \int r^{k}\av{n(r)}_{\Delta} dr$ is the
corresponding moment of the filtered DSD $\av{n(r)}_{\Delta}$ with
filter width $\Delta$. The fine-grained variance of $r$, we denote by $\sigma_{r}(0)$, is proportional to $\mu_{2}-\mu^{2}_{1}$ (up to a normalization constant), and $\sigma_{r}(\Delta) \propto \av{\mu_{2}}_{\Delta} - \av{\mu_{1}}^{2}_{\Delta}$ is the filtered variance of $r$.

{\color{black}{
We note that the moments $\mu_{k}$ and $\av{\mu_{k}}_{\Delta}$ are \textit{ensemble averages}, which are averages over realizations, the observation point being fixed. The spatial filtering specifies the sampling volume around the observation point. That is, the ensemble average we consider is \textit{conditional}. The ensemble average at a given observation point $\mbf x$ is computed over the droplet sizes \textit{conditional} on droplets positions being inside a given sampling \textit{volume} around $\mbf x$. This sampling volume can be: (a) $d \mbf x$ around $\mbf x$ (the ``physical point'' under the approximation of a continuous cloud) or (b) a finite volume $\Delta^{3}$ (the volume of the grid box) around a grid point $\mbf x$ in a simulation setting with spatial resolution $\Delta$. The conditional sampling in (a) corresponds to the fine-grained DSD, and the conditional sampling (b) corresponds to the filtered DSD with filter width $\Delta$ we discussed above.}}

Also, if $\Delta$ is large (compared to a lengthscale we shall
define later in Sec.~\ref{sec:cooper}), the ``grid point" may no
longer be considered mathematically infinitesimal as the ``physical
point", and the filtered DSD $\av{n(r)}_{\Delta}$ may incorporate
variation due to spatial distribution of the fine-grained DSD. Thus
fine-grained {\color{black}{DSDs}} that are locally narrow in a theoretical
physical point may produce filtered {\color{black}{DSDs}} that are locally
broad in a computational grid point.

{\color{black}{
\section{The reversible condensation regime}
\label{sec:cooper}

A useful theoretical framework to study the DSD formation was
developed by~\citet{Cooper_1989}. His analysis incorporates the
findings of earlier works [e.g., ~\citet{bartlett_jonas_qj_1972}
  and~\citet{Manton_qjrms_1979}] that rule out (under certain
conditions) turbulence-induced vertical velocity fluctuations as the
cause of strictly local DSD broadening. For clarity, we reproduce
below some essential points of~\citet{Cooper_1989} and make some
remarks. A general one is that theories attempting to explain DSD
broadening usually do not explicitly specify the sampling volume or
spatial averaging length scale $\Delta$ underlying the
examined DSD. We attempt to make this clear while recalling the
\textit{statistical ensemble} specified by~\citet{Cooper_1989}.

We follow~\citet{Cooper_1989} closely and assume that each droplet of
radius squared $a = r^{2}$ grows according to
\begin{equation}
\frac{da}{dt} = 2 D S(t),
\label{eq:growth_eq_a}
\end{equation}
where $S$ is the supersaturation experienced by the droplet and $D$ is the effective diffusion coefficient (here assumed constant for
simplicity). Surface tension and solute effects were neglected.

\citet{Cooper_1989} then proceeds by assuming the supersaturation
$S(t)$ has approached its quasi-steady value $S_{\text{qs}}$,
\begin{equation}
S \approx S_{\text{qs}} \equiv A_{1} \tau_{\text{ph}} w,
\label{eq:quasi-steady-approx}
\end{equation}
where $w$ is vertical velocity of the air parcel surrounding the
droplet ($A_{1}$ depends weakly on temperature and may be consider
constant to a good approximation), $\tau_{\text{ph}} = 1/(A_{2}
\mu_{1})$ is the phase relaxation time, $\mu_{1}$ is the first moment
of the DSD, and $A_{2}$ is a constant. $ \tau_{\text{ph}}$ is the time constant for the exponential approach of $S(t)$ to $S_{\text{qs}}$ and characterizes the local
structure of the cloudy environment where the droplet grows. Although~(\ref{eq:quasi-steady-approx})
is not valid everywhere (e.g., near the cloud base), the quasi-steady
regime is a useful limit for analysis.

The first moment $\mu_{1}$ specifying $\tau_{\text{ph}}$ can
be written as $\mu_{1} = N \av{r}$, where $N = \mu_{0}$ is the droplet number concentration (corresponding to the zeroth moment $\mu_{0}$ of
the DSD), and $\av{r} = \mu_{1}/N$ is the mean droplet radius. The sampling volume underlying $\mu_{1}$ and $\mu_{0}$ (and hence
$N$ and $\av{r}$) was not explicitly specified in~\citet{Cooper_1989},
thus we assume here that $\mu_{1}$ and $\mu_{0}$ are moments of the
fine-grained DSD defined in Sec.~\ref{sec:dsd}. 

The idealized models of turbulent condensation we will discuss here assume that 
\begin{equation}
\tau_{\text{ph}} \approx \bar{\tau}_{\text{ph}} \equiv 1/[A_{2}\av{\mu_{1}}_{\Delta}], 
\end{equation}
where $\Delta$ is the characteristic size of the sampling volume that include all droplets of the ensemble.

We also assume that the droplets are suspended and follow the air flow as tracers (droplet-tracer limit). Then the vertical component $z$ of the droplet position is related to the velocity $w$ by
\begin{equation}
dz = w(t) dt. 
\end{equation}
and hence
\begin{equation}
da \approx \bar{\gamma}_{a} dz, \; \; \; \; \bar{\gamma}_{a} \equiv 2DA_{1}\bar{\tau}_{\text{ph}}. 
\label{eq:da_dz_cooper}
\end{equation}

Now consider a droplet that at time $t_{0}$ has been activated with activation squared radius $a_{0}$ at the cloud base located at $z_{0}$. Under the assumptions above, the droplet squared radius at time $t > t_{0}$ for a droplet located at $Z(t) > z_{0}$ is
\[ a[Z(t)] \approx a_{\text{rev}}[Z(t)],\] where
\begin{equation}
a_{\text{rev}}(z) \equiv a_{0} + \bar{\gamma}_{a}(z - z_{0}),
\label{eq:a_z}
\end{equation}
is the \textit{reversible} squared radius. For fixed $a_{0}$ and $z_{0}$, Eq.~\eqref{eq:a_z} is a \textit{one-to-one relation} between the droplet squared radius $a$ and the droplet vertical position $z$. This defines the scenario of perfectly \textit{reversible
  condensation}\footnote{ {\color{black}{Consider a droplet of initial squared radius $a(t_{0})$ ascending with a time-dependent vertical velocity $w(t)$ for a certain time $t_{0} \leq t \leq t_{1}$. Now reverse the vertical velocity by setting $w(t) = - w[2(t_{1}-t_{0})-t]$ in $t_{1} \leq t \leq 2(t_{1}-t_{0})$. Under the reversible condensation regime, not only the droplet vertical motion will reverse, but its growth history reverses as well because $da \propto w(t) dt$. Hence its squared radius returns to its starting value $a(t_{0})$.}}}. 

    To discuss any characteristics of a droplet size distribution, we need to specify its underlying \textit{statistical
  ensemble}. The ensemble used by \citet{Cooper_1989} is comprised by all droplets reaching a given height $z$ at a particular time $t$
after being activated at time $t_{0}$ (with squared-radius $a_{0}$)
while crossing the cloud base located at
$z_{0}$. Also all droplet in the ensemble experiences the same constant phase relaxation time $\bar{\tau}_{\text{ph}}$. Thus for this ensemble, Eq.~(\ref{eq:a_z}) implies no variability in $a_{\text{rev}}$ at a particular height $z$
and the variance of $a_{\text{rev}}$ vanishes there,
\begin{equation}
\sigma^{2}_{a}(z,t) = \av{[a_{\text{rev}}-\av{a_{\text{rev}}}]^{2}} = 0.
\label{eq:var_a_def}
\end{equation}
We emphasize that the angle brackets $\av{\cdot}$
in~(\ref{eq:var_a_def}) denote an average over the ensemble of cloud droplets all having $Z(t) = z$ at time $t$ and having the same activation
area $a_{0}$ at the same activation level $z_{0}$. 

The scenario of \textit{reversible condensation} above is thus defined by three main assumptions: (a) quasi-steady supersaturation, (b) lack of
variability in the phase relaxation time $\tau_{\text{ph}}$ among the
droplets in the ensemble, and (c) the droplets are suspended and
follow the air flow as tracers (i.e., droplet inertia and
sedimentation are neglected). This third assumption is not made
explicit in the analysis of~\citet{Cooper_1989}.

This reversible condensation scenario will serve as a reference limit,
where {\color{black}{the droplet squared radius $a$ is \textit{maximally correlated} with its vertical position $z$ inside the cloud}}. Deviations from any of the assumptions (a) to (c)
listed above will work to destroy the one-to-one relation between $a$ and $z$,
thus creating conditions for broadening of the local (fine-grained)
DSD. 

\subsection{``Point" and extended ensembles}

\begin{figure*}[t]
    \centering
    \includegraphics[width=0.9\linewidth]{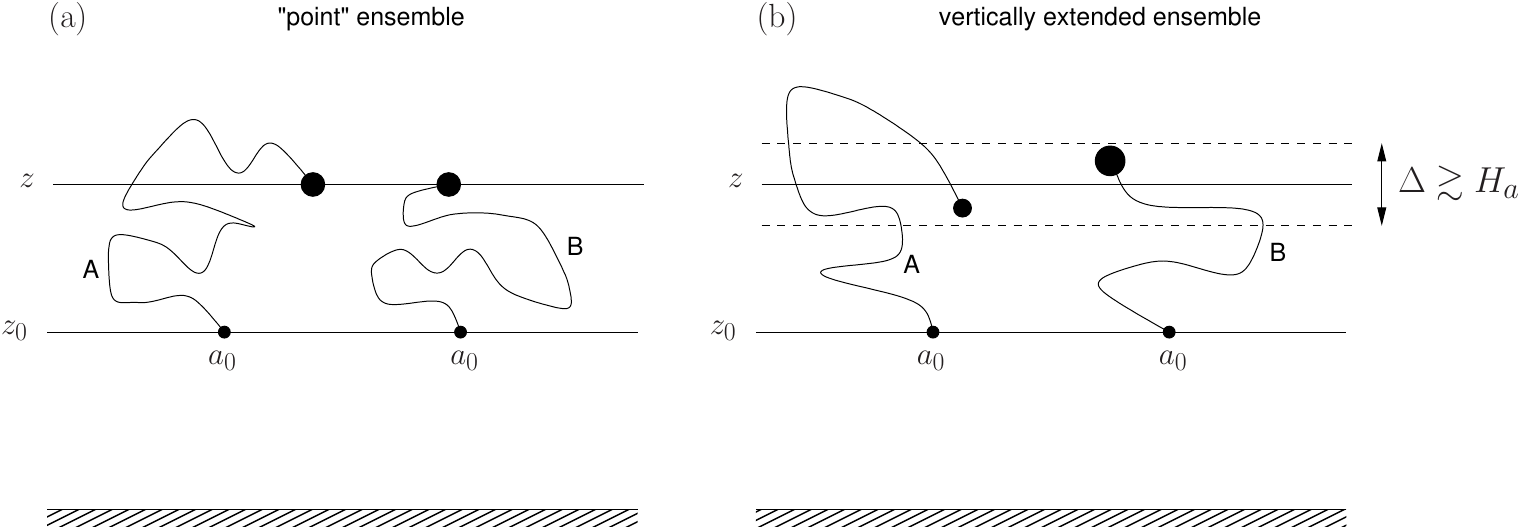}
    \caption{Schematic representation of (a) the ``point" ensemble
      ($\Delta \to 0$), and (b) the vertically extended
      ensemble (finite $\Delta$) {\color{black}{in an idealized cloud with reversible condensation}}. Two droplets undergoing
      distinct trajectories (labeled A and B) but arriving at the same
      particular height $z$ will have the same radius [panel (a)]. The breadth of the resulting fine-grained DSD vanishes. If
      the sampling volume is extended vertically to have a thickness
      $\Delta$, then at a given time $t$ this sampling
      volume may accommodate droplets arriving at different heights in
      the range $[z - \Delta/2, z + \Delta/2]$
      and thus having different radii [panel (b)]. The broad filtered DSD (with filter width $\Delta$) obtained under the reversible condensation regime [panel (b)] arises from the increase of droplet size
      with height [Eq.~(\ref{eq:a_z})] over the
      averaging length scale $\Delta$.}
    \label{fig:coopers_ensemble}
\end{figure*}

In the Cooper's ensemble discussed above, the ensemble averaging in Eq.~\eqref{eq:var_a_def} implicitly assumes $\Delta \to 0$ as the ensemble
considers droplets reaching a particular \textit{point} (at height
$z$) and forming the DSD there. Thus $\sigma^{2}_{a}(z,t)$ in Eq.~\eqref{eq:var_a_def} is the fine-grained variance $\sigma^{2}_{a}(0)$ at a ``physical point" in the
continuous cloud in the sense discussed in
Sec.~\ref{sec:dsd}. 

Figure~\ref{fig:coopers_ensemble}.a shows the
ensemble of droplets that reach a given height $z$ at a particular
time $t$. There, fluctuations in the updraft velocity only cannot
support broadening of the fine-grained DSD in the regime of reversible
condensation.

The situation is different in a simulation setting, where the sampling
volume is finite and defined by the length scale of the smallest
spatial variation resolved by the model. This motivates us to extend
Cooper's ``point" ensemble by considering the droplets that at time
$t$ reach a height in the range $[z - \Delta/2, z +
  \Delta/2]$, where $\Delta$ is of order of a
typical vertical length scale resolution used in LES. This ensemble is
illustrated in Fig.~\ref{fig:coopers_ensemble}.b. For this extended
ensemble, the broad filtered DSD (with filter width
$\Delta$) obtained in the reversible condensation scenario arises from the increase in droplet squared radius with
height over the averaging length scale $\Delta$ [Eq.~(\ref{eq:a_z})]. 

In principle, any particle-based SGS model for condensation built in
LES should capture this variability in the droplet squared
radius (at least under a regime of nearly reversible condensation). However, this variability is smeared out when the SGS microphysical scheme assumes all droplets in a given grid box experience the same
grid-mean supersaturation $\av{S}$, as this subgrid \textit{uniform
  condensation} tends to narrow the filtered DSD.

It is evident that the breadth of the local filtered DSD will depend on $\Delta$ when $\Delta$ is of the same
order or larger than the length scale 
\begin{equation}
H_{a} \equiv \sigma_{a}(0)/|\partial a_{\text{rev}}| = \sigma_{a}(0)/\bar{\gamma}_{a},
\label{eq:scale_height_a}
\end{equation}
where $\sigma_{a}(0)$ denotes the \textit{fine-grained} standard deviation of the droplet squared radius (corresponding to a vanishing sampling volume or filter width $\Delta \to 0$). Then $H_{a}$ is the length scale over which the reversible squared radius $a_{\text{rev}}$ changes by a value $\sim \sigma_{a}(0)$ along the vertical direction. Thus for $\Delta \gtrsim H_{a}$, the filtered DSD with filter width $\Delta$ will incorporate variation due to spatial distribution of the fine-grained DSD and $\sigma_{a}(\Delta)$ will appreciably differ from $\sigma_{a}(0)$.

Later in this work we use Cooper's framework and the reference
reversible condensation scenario outlined above to discuss the current
limitations of two idealized frameworks, namely the so-called eddy
hopping model operating in adiabatic parcels and certain DNS-like
models, where microphysical variability is {\color{black}{driven}} by local velocity
fluctuations.
}}

\section{Models operating in adiabatic cloud parcels}
\label{sec:parcel}

The standard theory of droplet growth by condensation in an adiabatic
cloud parcel assumes uniform condensation, where all droplets in the
parcel experience the same supersaturation. This approximation
produces narrow DSDs in disagreement with
\textit{in situ} measurements of broad droplet spectra in clouds [see e.g.,~\citet{Shaw_arfm_2003}].

Attempts to resolve this paradox attributes the occurrence of broad
droplet size spectra in clouds to the stochastic droplet growth by
condensation in a turbulent environment [see
  e.g.,~\citet{KC_1999a}]. An approximate account of turbulent effects
on condensation in the adiabatic cloud parcel framework was done
by~\citet{Sardina_et_al_2015},~\citet{Grabowski_Abade_2017}~and~\citet{Abade_et_al_2018}. These so-called eddy
hopping models use a Lagrangian (particle-based) stochastic treatment
of the droplet growth and presumably predict a broad DSD. 

A much earlier numerical study by~\citet{bartlett_jonas_qj_1972}
attempted to account for the effects of turbulence by subjecting the
adiabatic cloud parcel to a time-dependent updraft. Their study ruled out turbulence-induced vertical velocity fluctuations as the cause of strictly local DSD broadening. Then why does the eddy hopping model~\citep{Sardina_et_al_2015,Grabowski_Abade_2017,Abade_et_al_2018} predict broad DSDs,
while~\citet{bartlett_jonas_qj_1972} do not? 

The essential difference
between these two approaches lies in the statistical ensemble over
which the DSD is sampled. The ensemble
in~\citet{bartlett_jonas_qj_1972} corresponds to the ``point" ensemble (Fig.~\ref{fig:coopers_ensemble}a)
with infinitesimal sampling length scale ($\Delta \to 0$),
and thus provides the \textit{fine-grained} DSD. In contrast, the eddy hopping
model considers the extended ensemble (Fig.~\ref{fig:coopers_ensemble}b) with finite
sampling length scale, and thus provides the \textit{filtered} DSD with filter
width $\Delta \gtrsim H_{a}$. While the fine-grained DSD
of~\citet{bartlett_jonas_qj_1972} is narrow, the filtered DSD predicted by the eddy hopping model is broad. 

Next we discuss the eddy hopping model in more detail and explain a major
artifact in~\citet{Sardina_et_al_2015,Grabowski_Abade_2017,Abade_et_al_2018}. This artifact makes
the variance of the filtered DSD in the adiabatic parcel to
continuously grow in time. We show that this growth does not reflect
an actual local DSD broadening with time but results from the growth
in time of the filter width $\Delta$ underlying the
examined DSD.

{\color{black}{

\subsection{Eddy hopping model and nonlocal DSD sampling}

The eddy hopping model considers an ensemble of representative cloud droplets. The model describes the evolution of four variables attached to each droplet: its squared radius $a$, the supersaturation $S$ the droplet experiences, its vertical position $z$ and its vertical velocity $w$. The evolution of the state variables $(a,S,w,z)$ is described by the following set of equations:
\begin{equation}
    \frac{da}{dt} = 2 D S,
    \label{eq:sde_aprime}
\end{equation}
\begin{equation}
    \frac{d S}{d t} = - \frac{S}{\tau_{s}} + A_{1} w, 
    \label{eq:sde_sprime}
\end{equation}
\begin{equation}
d w = - \frac{w}{\tau_{w}}  d t + \sqrt{\frac{2 \sigma^{2}_{w}}{\tau_{w}} } d W(t),  
\label{eq:dw_prime}
\end{equation}
\begin{equation}
d z = w d t.   
\label{eq:dz_prime}
\end{equation}
The equations above include a set of model parameters $(\tau_{s},\tau_{w},\sigma_{w})$ we shall specify later. 

The identifying feature of the eddy hopping model is that the supersaturation evolving according to~\eqref{eq:sde_sprime} is forced  by vertical velocity $w$ via the adiabatic cooling of the ascending air surrounding the cloud droplet. The vertical velocity $w$ is described by Eq.~\eqref{eq:dw_prime}, where $\tau_{w}$ is the Lagrangian integral timescale of $w$, and $dW(t)$ is the increment of a Wiener process [see e.g.,~\citet{Rodean_book}]. The latter introduces the ``noise" due to turbulent fluctuations.  

Equation~\eqref{eq:dw_prime} serves as a coarse representation of the vertical velocity fluctuations of a fluid particle due to a range of homogeneous isotropic turbulence eddies of sizes smaller than a characteristic length $\Delta$. This is the length scale on which the cloud parcel, filled with homogeneous turbulence, is imagined to acquire its
mean turbulent kinetic energy $k$ $(= 3\sigma^{2}_{w}/2)$ at a given
rate $\varepsilon$ per unit mass. 

Assuming that turbulent cloud parcels may provide insights into subgrid modeling of the microphysics in large-eddy simulations (LES) of clouds, it is
reasonable to assume that $\Delta$ has the size of a typical LES grid box. Thus $\Delta$ lies in the inertial range of cloud turbulence and
usually varies from several meters to several tens of meters.

Equation~(\ref{eq:sde_sprime}) for the supersaturation may be cast into the form
\begin{equation}
\frac{d S}{d t} = - \frac{S- S_{\text{qs}}}{\tau_{s}},
\label{eq:sde_sprime_relax}
\end{equation}
to make it explicit that $S$ relaxes to the quasi-steady value
$S_{\text{qs}} = A_{1} \tau_{s} w$ with time constant $\tau_{s}$ --- the supersaturation relaxation time. The time constant $\tau_{s}$ is a model parameter defined as the harmonic mean of the phase relaxation time $\tau_{\text{ph}}$ and the relaxation time $\tau_{\text{mix}}$ associated with turbulent
mixing, $\tau_{s} = 1/(\tau^{-1}_{\text{ph}} + \tau^{-1}_{\text{mix}})$. We commonly assume $\tau_{\text{mix}} \approx \tau_{w}$.  

For times sufficiently longer than $\tau_{s}$, the supersaturation has attained its quasi-steady value $S \simeq S_{\text{qs}}$ which is proportional to $w(t)$ [as in Eq.~\eqref{eq:quasi-steady-approx}]. Then it follows from~\eqref{eq:dz_prime} that the increment $da$ in the droplet squared radius is closely related to the increment $dz$ in its vertical position, 
\begin{equation}
da \simeq \bar{\gamma}_{a} dz,
\label{eq:da_propto_dz}
\end{equation}
where $\bar{\gamma}_{a} = 2DA_{1} \tau_{s}$ and we have assumed that $\tau_{s}$ is the same for all representative droplets in the ensemble.  

Also, it follows from Eqs.~\eqref{eq:dw_prime} and~\eqref{eq:dz_prime} that (for times sufficiently longer than $\tau_{w}$) the mean squared displacement along the vertical direction grows like [see e.g.,~\citet{Lemons}]
\begin{equation}
\av{[z(t)-\av{z}]^{2}} \sim t,
\label{eq:msd_z}
\end{equation}
being the average $\av{\cdots}$ calculated over all the representative cloud droplets. Thus the relation~\eqref{eq:da_propto_dz} implies that the variance $\sigma^{2}_{a}$ of droplet squared radius grows according to the same scaling,   
\begin{equation} 
\sigma^{2}_{a} = \av{[a(t)-\av{a}]^{2}} \sim t,
\label{eq:var_a_scaling}
\end{equation}
when the average is computed over the same ensemble of representative droplets. 

In conclusion, the eddy hopping model predicts continuous DSD broadening as the variance $\sigma^{2}_{a}$ grows with time~\citep{Sardina_et_al_2015}. However, this broadening is \textit{non-local} in a sense to be explained soon below. This non-local feature underlying the continuous growth of $\sigma_{a}$ was overlooked by~\citet{Sardina_et_al_2015,Grabowski_Abade_2017,Abade_et_al_2018} because they
inadvertently omitted Eq.~(\ref{eq:dz_prime}) from their description. 

\begin{figure*}
    \centering
    \includegraphics[width=0.8\linewidth]{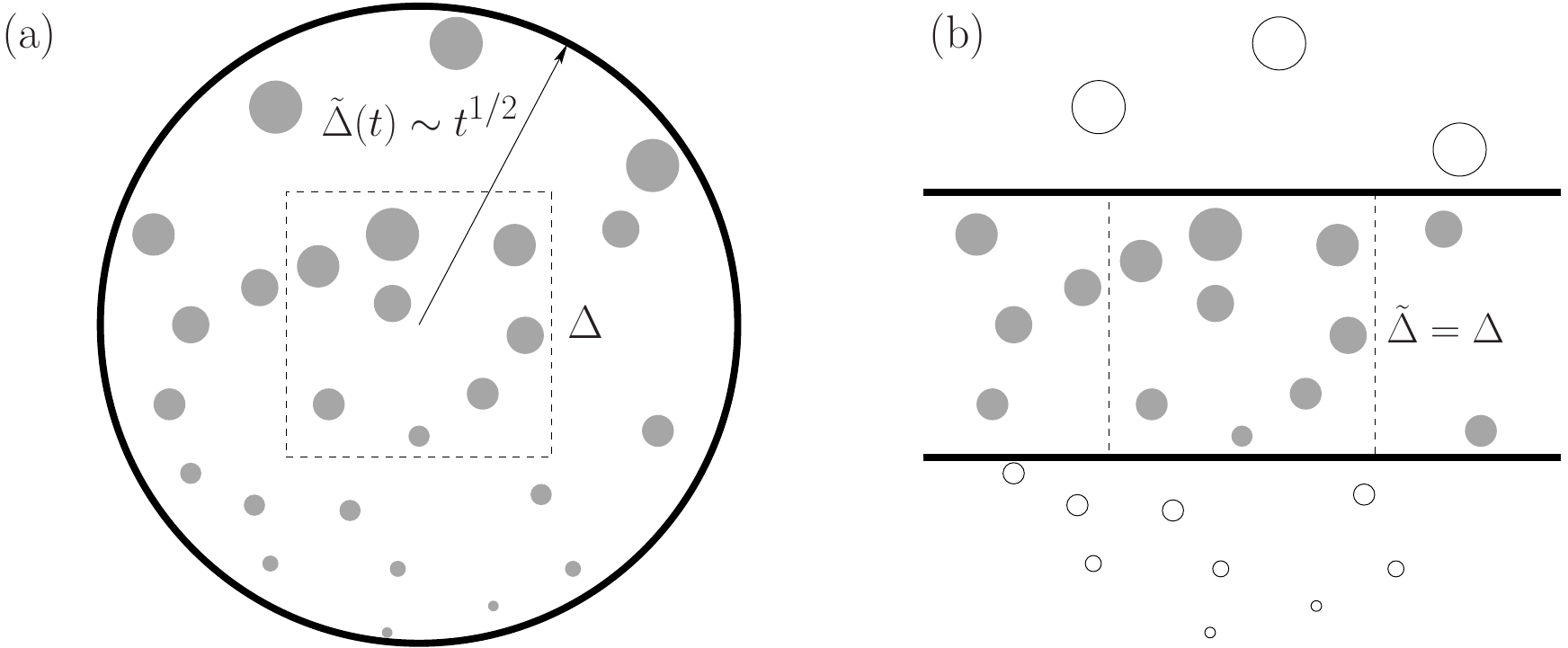}
    \caption{(a) Non-local sampling in a turbulent adiabatic parcel,
      where the linear extent $\tilde{\Delta}(t)$ of the implicit sampling volume growths like $\sim t^{1/2}$ due to turbulent diffusion of
      cloud particles in all directions. As time increases,
      $\tilde{\Delta}(t)$ becomes larger than the characteristic
      length scale $\Delta$ on which the parcel acquires its turbulent
      kinetic energy. The same problem occurs in DNS-like models with usual periodic boundary conditions (PBC) in the vertical direction (see discussion in
      Sec.~\ref{sec:dns}). (b) Adiabatic parcel with DSD statistics sampled over a subensemble of droplets (in gray) using proper boundary conditions at the top and bottom to enforce
      local sampling (see text for details).}
    \label{fig:parcel_boundary_conditions}
\end{figure*}

The fluctuating vertical
velocity $w$ that drives supersaturation fluctuations $S$ also drives the vertical dispersion of cloud particles. The spatial extent of this droplet dispersion is measured by the mean squared displacement~(\ref{eq:msd_z}). Thus the associated length scale
\begin{equation}
\tilde{\Delta}(t) =  \av{[z(t)-\av{z}]^{2}}^{1/2} \sim t^{1/2},
\label{eq:delta_av_growth_time}
\end{equation}
which may be understood as the linear size of the sampling volume containing the whole droplet population in the parcel, grows with
the square root of time. The sampling length scale $\tilde{\Delta}(t)$ is not bounded and may eventually exceed the fixed scale $\Delta$ on which energy is assumed to be supplied to the turbulence filling
the parcel. We refer to this as a \textit{non-local sampling} of the DSD in the eddy hopping model as illustrated schematically in
Fig.~\ref{fig:parcel_boundary_conditions}.a. Figure~\ref{fig:scatter_reversible} shows numerical results for the continuous droplet dispersion along the vertical extrapolating $\Delta$ [gray dots in panels (a) and (b)] and the corresponding continuous growth of the standard deviation $\sigma_{a}[\tilde{\Delta}(t)]$ [gray line in panel (c)].  

\begin{figure*}
    \centering
    \includegraphics[width=1.0\linewidth]{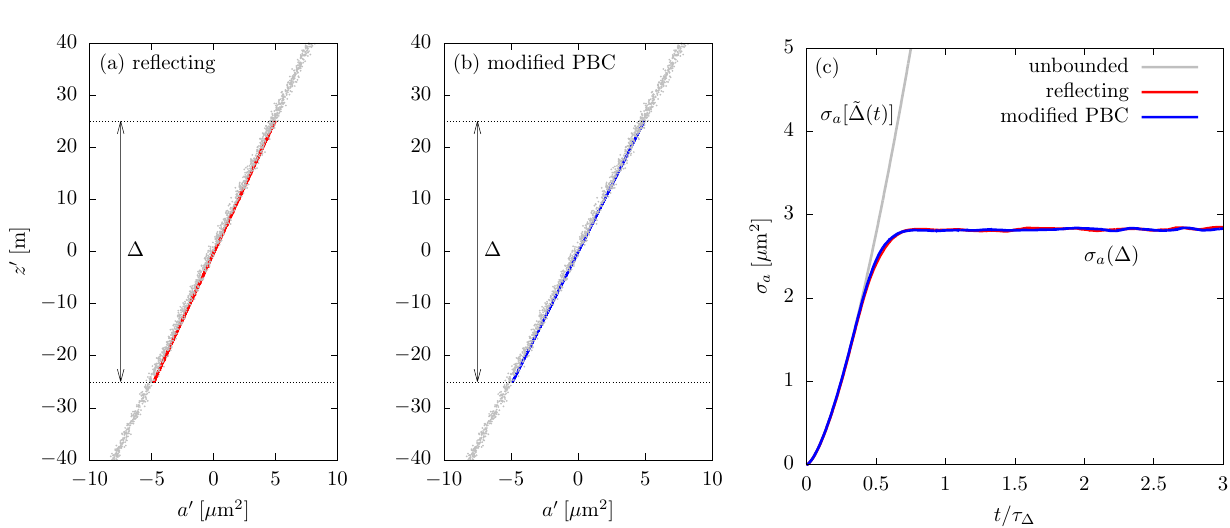}
    \caption{[Panels (a) and (b)] Scatterplots of $a'= a-\av{a}$ and $z' = z - \av{z}$ to illustrate the local sampling in an adiabatic parcel of size $\Delta$ enforced by using (a) reflecting boundary conditions (red dots) and (b) modified periodic boundary conditions (blue dots). Gray dots in panels (a) and (b) are for an unbounded sampling volume (non-local sampling) of growing length $\tilde{\Delta}(t) \sim t^{1/2}$. The results consider the reversible condensation regime. Panel (c) shows the time evolution $(\tau_{\Delta} = \varepsilon^{-1/3}\Delta^{2/3})$ of the filtered standard deviation $\sigma_{a}(\Delta)$ for fixed filter width $\Delta$ (reflecting top and bottom walls and modified PBC), while $\sigma_{a}[\tilde{\Delta}(t)]$ continuously grows along with the sampling length scale $\tilde{\Delta}(t)$ in the unbounded case. Simulations are for $N = 100 \: \text{cm}^{-3}$, $\varepsilon = 10^{-3} \: \text{m}^{2} \text{s}^{-3}$, $\av{S} = 0$, and $\av{a} = a_{0} = 100 \: \mu\text{m}^{2}$.}
    \label{fig:scatter_reversible}
\end{figure*}

\subsection{Enforcing local sampling of the DSD}
\label{sec:spurious_broadening}

For the eddy hopping model to be relevant as an SGS stochastic condensation scheme in LES, then $\tilde{\Delta}$ (the linear extent of the volume over which droplet size statistics is sampled) should be bounded to $\Delta$. This ensures a well-defined filtered variance $\sigma^{2}_{a}(\Delta)$ that refers to the droplet population inside a grid box of size $\Delta$. This motivates us to enforce local sampling of the DSD in the adiabatic parcel and DNS-like frameworks as well. 

We consider three different ways of imposing local DSD sampling depending on the framework where the eddy hopping model operates. 

First, we may sample the droplet size statistics over a \textit{subensemble} of cloud particles all having vertical position $Z(t)$ between $\av{z} - \Delta/2$ and $\av{z} - \Delta/2$, as illustrated in Fig.~\ref{fig:parcel_boundary_conditions}.b. This approach can be used in LES (as we shall discuss in Sec.~\ref{sec:sgs_model}), but is it impractical in the adiabatic parcel and DNS-like frameworks, because the number of droplets in the subensemble decays with time as droplets disperse and leave the sampling volume. 
 
A second approach for local sampling is by using reflecting boundary conditions on $z$ and $w$ at the bottom and top horizontal boundaries of the sampling volume. These boundaries are separated by a constant
distance equal to $\Delta$, thus enforcing $\tilde{\Delta} = \Delta$ as illustrated in Fig.~\ref{fig:parcel_boundary_conditions}.b. This treatment was used by~\citet{Abade_Albuquerque_2024} in an extended version of the eddy hopping model operating in mixed-phase adiabatic parcels. Reflecting boundary conditions preserve the number of representative droplets in the sampling volume. However it can be used only in the adiabatic parcel framework under the reversible condensation regime.  

\begin{figure}
    \centering \includegraphics[width=1.0\linewidth]{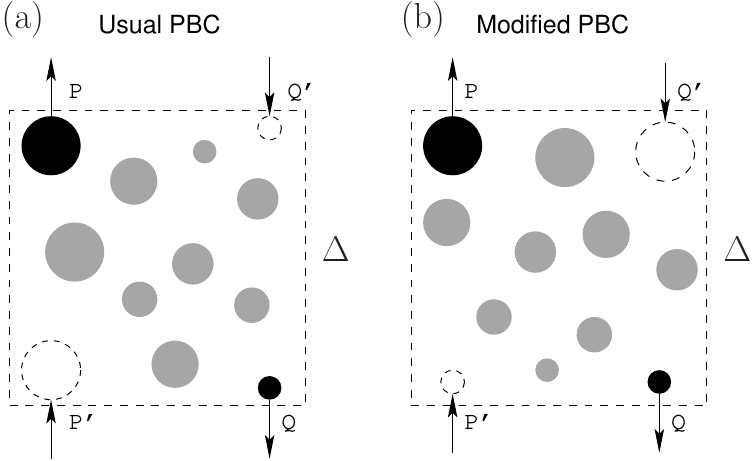}
    \caption{(a) Under usual PBC a large droplet leaves the sampling box at some point $P$ and is reintroduced into the box at $P'$ with its radius unaltered. Accordingly, a small droplet that leaves the box at $Q$ is reintroduced at $Q'$ with the same radius. (b) The modified PBC shrinks the radius of a large droplet as it is reintroduced at $P'$ after leaving the box at the point $P$. Accordingly, modified PBC expands the radius of a small droplet as it is reintroduced at $Q'$ after leaving the box at the point~$Q$.}
    \label{fig:usual_vs_modified_pbc}
\end{figure}

A third approach is the use of so-called \textit{modified} periodic boundary conditions (PBC) in the vertical direction to be explained below. As Equation~\eqref{eq:da_propto_dz} and Figures~\ref{fig:parcel_boundary_conditions},~\ref{fig:scatter_reversible}.a,~\ref{fig:scatter_reversible}.b indicate, the sampling volume containing droplets is embedded in a cloud which has a mean squared radius gradient
$\bar{\gamma}_{a} \: (\approx \partial_{z}a_{\text{rev}} = 2DA_{1} \tau_{s}$) in the $z$-direction. This resembles non-equilibrium (particle-based) molecular dynamics simulations of a fluid under shear~\citep{Lees_Edwards_1972} with constant velocity gradient $\gamma_{u} = du/dz$ in the $z$-direction\footnote{In molecular dynamics simulations of fluid under shear, the sampling volume containing the particles is considered as being embedded in a fluid which has a constant velocity gradient $\gamma_{u} = du/dz$ in the $z$-direction.}. In this case, PBC must be used not only on particle positions, but on particle velocities as well~\citep{Lees_Edwards_1972}. 

In our turbulent condensation problem, modified PBC (in contrast to \textit{usual} PBC) remaps not only droplets vertical positions, but also adjusts the droplets radii as droplets cross the the bottom and top horizontal boundaries of the sampling volume. Usual and modified PBC are contrasted in Fig.~\ref{fig:usual_vs_modified_pbc}. Imagine that a droplet leaves the sampling volume at the top boundary [point $P$ in panels (a) and (b) of Fig.~\ref{fig:usual_vs_modified_pbc}] with large squared radius. Under \textit{usual} periodic boundary conditions the droplet will be reintroduced into the sampling volume at the bottom boundary (point $P'$ of Fig.~\ref{fig:usual_vs_modified_pbc}.a) with its large squared radius unaltered. However, due to the gradient $\bar{\gamma}_{a}$, the squared radius must be also adjusted when the droplet is reintroduced at point $P'$ (as in Fig.~\ref{fig:usual_vs_modified_pbc}.b) to produce a stable squared radius gradient over sufficiently long periods in the steady state. The modified PBC scheme is described in detail in Appendix~A. 

As for reflecting boundaries~\citep{Abade_Albuquerque_2024}, modified PBC preserve the number of representative droplets in the ensemble. As an advantage, modified PBC can be (presumably) applied in DNS-like studies as well in the droplet-tracer limit (sedimentation and droplet inertia neglected). Also, modified PBC can be used to extract the fine-grained DSD due to deviations from the reversible condensation regime, as we shall discuss in Sec.~\ref{sec:clark_hall}.

Figure~\ref{fig:scatter_reversible} shows that reflecting boundaries and the modified PBC are effective to enforce local sampling of the DSD and produce the same well-defined filtered variance $\sigma_{a}(\Delta)$ for filter width $\Delta$.

}}

{\color{black}{
\subsection{A reduced stochastic model}
\label{sec:reduced_stochastic_model}

It follows from previous sections that the eddy hopping model builds
up correlations between droplet sizes and droplet vertical
positions. The resulting fine-grained DSDs are spatially inhomogeneous
(along the vertical direction) and in general differ from the
$\Delta$-dependent filtered DSD. It is then instructive to consider a
simple model of stochastic condensation [as the one discussed
  by~\citet{Chandrakar_et_al_2016} and~\citet{Saito_et_al_2019}] that
coarsely represents DSD broadening in an idealized scenario, where the
statistics of the droplet sizes is imagined to be spatially
homogeneous. This scenario emerges in a certain type of DNS-like
models of turbulent condensation that mimics an idealized homogeneous
and unbounded cloud (to be discussed later in Sec.~\ref{sec:dns}).

The reduced model comprises two equations, one for $a$ and other for $S$,
\begin{equation}
    da = 2D S dt,
    \label{eq:sde_aprime_saito}
\end{equation}
\begin{equation}
dS = - \frac{S}{\tau_{s}} d t + \sqrt{\frac{2 \sigma^{2}_{s}}{\tau_{s}}} \: dW(t). 
\label{eq:sde_sprime_saito}
\end{equation}
Unlike the eddy hopping model, the reduced model describes only two
jointly distributed stochastic variables, $a$ and $S$, regardless of
the droplets positions and velocities. Also, the ``noise" representing
fluctuations enters directly in the Eq.~\eqref{eq:sde_aprime_saito}
for supersaturation. Here, we assume the model parameters $\sigma_{s}$
and $\tau_{s}$ in Eq.~\eqref{eq:sde_aprime_saito} can be
estimated. For the moment, we leave unspecified the physical processes
that produce supersaturation fluctuations (and determine $\sigma_{s}$
and $\tau_{s}$).

It can be shown [see e.g.,~\citet{Chandrakar_et_al_2016}] that this
reduced model produces the same asymptotic continuous
growth~(\ref{eq:var_a_scaling}) of the variance of $a$ (for $t$
sufficiently longer than $\tau_{s}$),
\begin{equation}
\sigma^{2}_{a} = K \: t, \; \; \; \; t \gg \tau_{s},
\label{eq:sigma_a_scaling_simplified}
\end{equation}
where $K \equiv 8 D^{2} \sigma^{2}_{s} \tau_{s}$. However, if
fluctuations in $S$ are created by sources other than vertical
velocity fluctuations, then the continuous growth of $\sigma^{2}_{a}$
in~\eqref{eq:sigma_a_scaling_simplified} does not have the spatial
interpretation as in~(\ref{eq:var_a_scaling}) because $a$ and $S$ and
are decoupled from the droplet transport. The resulting fine-grained
DSD is \textit{spatially homogeneous} and thus coincides with the
filtered DSD for arbitrary filter width $\Delta$.

The underlying sources of supersaturation fluctuations in the reduced
model (as those due to vertical velocity fluctuations are excluded)
may be stationary external boundary conditions that sustain the scalar
fluctuations against turbulent mixing and scalar
dissipation\footnote{In the Pi Chamber, stationary conditions at the
top, bottom, and sidewalls determine how scalar fluctuations are
maintained against turbulent mixing and scalar dissipation. Also, as
the characteristic vertical displacement of air parcels in the Pi
Chamber is a negligible fraction of the length scale $1/A_{1}$,
different adiabatic cooling rates arising from turbulent fluctuations
in the vertical velocities cannot be regarded as a source of
variability in droplet growth conditions.}. This external boundary
condition determine the supersaturation variance $\sigma^{2}_{s}$, a
model parameter. The other model parameter is $\tau_{s}$, the
Lagrangian integral scale of $S$ [Eq.~\eqref{eq:tau_s_def}], which is
difficult to measure in Lagrangian coordinates. Then $\tau_{s}$ can be
estimated by assuming the second-order Lagrangian structure function
of $S$, $\av{dS^{2}(\tau)} \equiv \av{[S(t+\tau) - S(t)]^{2}}$,
follows the Kolmogorov scaling,
\begin{equation}
\av{dS^{2}(\tau)} = C_{1} \chi_{s} \tau,
\label{eq:koc}
\end{equation}
where $C_{1}$ is a constant, $\chi_{s}$ is the mean dissipation rate of the scalar variance $\av{S'^{2}}$, and the time lag $\tau$ lies in the inertial range (i.e., $\tau_{\eta} \ll s \ll \tau_{s}$, where $\tau_{\eta}$ is the Kolmogorov timescale). Relations of type~\eqref{eq:koc} were thoroughly tested by~\citet{Chandrakar_etal_jas_2023} in high-resolution simulations of the Pi Chamber setup. 

Then squaring Eq.~\eqref{eq:sde_sprime_saito}, taking the ensemble
average, and using~\eqref{eq:koc} yield the amplitude $\sqrt{C_{1}
  \chi_{s}} \: ( = \sqrt{2 \sigma^{2}_{s}/\tau_{s}})$ of the random
term in Eq.~\eqref{eq:sde_sprime_saito} and the time scale $\tau_{s} =
2 \sigma^{2}_{s}/(C_{1} \chi_{s})$. The model parameters ($\sigma_{s},
\tau_{s}$) here have no correlation with the statistics of droplets
vertical velocities.
 
The four-equation eddy hopping model (Sec.~\ref{sec:parcel}) and the
two-equation reduced stochastic model described above serve as
prototypes of the two classes of DNS-like models we shall discuss
later in Sec.~\ref{sec:dns}.

}}

{\color{black}{

\subsection{The incomplete equivalence between reduced and eddy-hopping models}
\label{sec:incomplete_equivalence}

The scaling $\sigma^{2}_{a} \sim t$ [Eqs.~\eqref{eq:var_a_scaling} and~\eqref{eq:sigma_a_scaling_simplified}] for the variance is robust and will be observed in both eddy hopping and reduced stochastic models. As $da/dt \propto S(t)$ (a linear relation) and $S(t)$ is a stochastic process, then $\sigma^{2}_{a}$ will eventually grow like $t$ for times sufficiently longer than the supersaturation autocorrelation time $\tau_{s}$, defined as 
\begin{equation}
\tau_{s} = \frac{1}{\sigma^{2}_{s}} \int_{0}^{\infty} R_{s}(t) dt,
\label{eq:tau_s_def}
\end{equation}
where $R_{s}(t) = \av{S'(t_{0})S'(t_{0}+t)}$ is the supersaturation autocorrelation function, with $S'= S - \av{S}$ and $\sigma^{2}_{s} = R_{s}(0)$. 

However, the essential point here is the \textit{source} of the random fluctuations of $S(t)$. The DSD broadening described by the reduced model [Eqs.~\eqref{eq:sde_aprime_saito} and~\eqref{eq:sde_sprime_saito}] can be regarded as \textit{strictly local} only if the noise in~\eqref{eq:sde_sprime_saito}, which is parametrized by $\sigma_{s}$ and $\tau_{s}$, is due to sources other than vertical velocity fluctuations. If the parametrization of supersaturation fluctuations include any source due to vertical velocity fluctuations, then this forcing will create correlations between droplets radii and droplet vertical positions. This implies that the fine-grained and filtered DSDs no longer coincide and the filtered DSD will depend on the filter width $\Delta$. 

It has been suggested~\citep{Saito_et_al_2021} that the two-equation model above~[(\ref{eq:sde_aprime_saito})
  and~(\ref{eq:sde_sprime_saito})] can be obtained by simplifying the
eddy hopping model. \citet{Saito_et_al_2021}
showed that in the large scale limit, where the microphysics is fast
compared to the slow turbulent mixing in large systems
($\tau_{\text{ph}} \ll \tau_{\text{mix}} \sim \Delta^{2/3}$), the
vertical velocity fluctuation $w$ can be eliminated from the
stochastic eddy hopping model. As shown by~\citet{Saito_et_al_2021},
the statistical properties of supersaturation fluctuations in the simplified model (namely, $\sigma_{s}$, $R_{s}(t)$, and thus $\tau_{s}$) converge to those
of the model without simplification as the ratio
$\tau_{\text{mix}}/\tau_{\text{ph}} \gg 1$ increases (i.e., as one
approaches the large scale limit).

However, the elimination of $w$ from
the description masks the issues of non-local sampling discussed in
Sec.~\ref{sec:parcel}.\ref{sec:spurious_broadening}. Also it removes the possibility of enforcing local sampling by using the approaches discussed in Sec.~\ref{sec:parcel}.\ref{sec:spurious_broadening}.   

Moreover, although $w$ is eliminated in the
simplified model of~\citet{Saito_et_al_2021}, vertical velocity fluctuations are the hidden source of supersaturation fluctuations as $\sigma_{w}$ and $\tau_{w}$ remain in the expressions for the parameters $\sigma_{s}$ and $\tau_{s}$ [see Eqs.~(22) and~(30) in~\citet{Saito_et_al_2021}]. At first sight, this procedure may suggest a ``statistical equivalence" between the eddy-hopping model and its reduced version. However, this equivalence is incomplete, as the reduction procedure fixes only $\sigma_{s}$ and $R_{s}(t)$, but hides correlations between droplets sizes and droplets vertical positions.

In an LES setting with
particle-based microphysics, a proper implementation of the eddy
hopping as an SGS microphysical model requires coupling between the
vertical SGS droplet transport and the vertical velocity fluctuation
$w$ that drives the SGS supersaturation fluctuations, as we shall
discuss in Sec.~\ref{sec:sgs_model}. Then the vertical velocity fluctuation should remain as an explicit model variable.
}}

\section{Deviations from the quasi-steady supersaturation}
\label{sec:clark_hall}

As discussed in Sec.~\ref{sec:cooper}, deviations
from any of the three main assumptions [from (a) to (c)] underlying
the reversible condensation scenario will work to destroy correlations
between droplet size and droplet vertical position, creating
conditions for strictly local broadening of the fine-grained DSD. The eddy hopping model incorporates two of these assumptions,
namely, that all droplets in the ensemble experience the same
supersaturation relaxation time $\tau_{s}$ and that droplets follow the air flow as tracers (sedimentation and inertia are
neglected). The eddy hopping model then allows for deviations from
the quasi-steady supersaturation [Eq.~(\ref{eq:quasi-steady-approx})], being these deviations the only possible
cause of broadening of the fine-grained DSD in the adiabatic
(non-entraining) parcel framework.

{\color{black}{
In general, the supersaturation driven by vertical velocity
fluctuations is frequency-dependent, that is, its response to
fluctuations in the vertical velocity is retarded. This is clear after
further inspection of Eqs.~\eqref{eq:sde_sprime}  and~\eqref{eq:sde_sprime_relax}, which gives the
linear relation
\begin{equation}
S(\omega) = \varphi(\omega) w(\omega), \; \; \; \; \; \; \; \varphi(\omega) = \frac{A_{1} \tau_{s}}{ 1 - i \omega \tau_{s} },
\label{eq:s_frequency}
\end{equation}
between the Fourier-Laplace transforms $S(\omega)$ and
$w(\omega)$ of $S(t)$ and $w(t)$,
respectively, where $\omega$ is the frequency, and $\varphi(\omega) = \varphi'(\omega) + i \varphi'' (\omega) = |\varphi|e^{i\theta}$ is the amplitude- and phase-modifying transfer function. The phase modification $\theta = \arctan(\varphi''/\varphi')$ is due to the generally non-zero imaginary part $\varphi'' (\omega)$. More details in Appendix~B.

The quasi-steady
approximation corresponds to the low-frequency limit $\omega \tau_{s}
\to 0$, where $S(\omega)$ and $w(\omega)$ [and thus $S_{\text{qs}}(\omega)
\equiv A_{1} \tau_{s} w(\omega)$] are in phase (as $\varphi''$ vanishes). For finite and high frequencies $\omega \tau_{s}$, $S(\omega)$ and $w(\omega)$ are phase-shifted. Then the close
       link between supersaturation and vertical velocity is broken [see e.g.,~\citet{KC_1999a}]
       and the condensation is no longer perfectly reversible (even
       when there is no variability in $\tau_{s}$ among the droplets
       of the ensemble).
}}

The quasi-steady approximation~(\ref{eq:quasi-steady-approx}) cannot represent {\color{black}{high-frequency supersaturation fluctuations. That is, fluctuations}} occurring on time scales
shorter than the time constant $\tau_{\text{s}}$ characterizing the
relaxation of $S$ to the quasi-steady value $S_{\text{qs}}$
[Eq.~(\ref{eq:sde_sprime_relax})]. Associated with $\tau_{s}$ there is a spatial lengthscale
$\ell_{\text{s}}$ (the supersaturation relaxation length). It can be estimated as~\citep{Kabanov_etal_jas_1971,Clark_and_Hall_1979},
\begin{equation}
\ell_{\text{s}} \sim \varepsilon^{1/2} \tau_{\text{s}}^{3/2},
\label{eq:ell_s}
\end{equation}
where $\varepsilon$ is the mean dissipation rate. The
relation~(\ref{eq:ell_s}) is obtained on dimensional grounds by
assuming that both $\tau_{\text{s}}$ and $\ell_{\text{s}}$ lie in the
Kolmogorov's inertial range of cloud turbulence. This allows us to
express the range of description supported by the quasi-steady
approximation in terms of spatial lengthscales: the quasi-steady
supersaturation fluctuation $S_{\text{qs}}$ cannot represent
microphysical variability due to vertical velocities of eddies of
spatial scales smaller than $\ell_{\text{s}}$.

\begin{figure}
    \centering
    \includegraphics[width=1.0\linewidth]{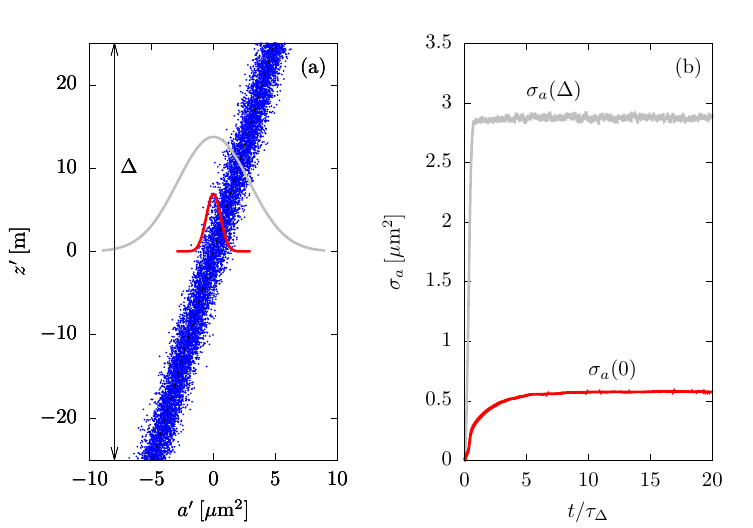}
    \caption{Fine-grained \textit{versus} filtered DSD for finite $\tau_{s}$. (a) Scatterplot of $z'$ and $a'$ (blue dots) along with the broad filtered DSD of filter width $\Delta = 50 \: \text{m}$ (gray line) and the much narrower fine grained DSD of \textit{vanishing} filter width (red line); (b) time evolution of the filtered $\sigma_{a}(\Delta)$ and fine-grained $\sigma_{a}(0)$ standard deviation of $a$. Simulations using the eddy hopping model with modified PBC are for $N = 100 \: \text{cm}^{-3}$, $\varepsilon = 10^{-3} \: \text{m}^{2} \text{s}^{-3}$, $\Delta = 50 \: \text{m}$, $\av{S} = 0$, and $\av{a} = a_{0} = 100 \: \mu\text{m}^{2}$. The scale height is $H_{a} \approx 3 \: \text{m}$ in the steady-state.}
    \label{fig:fine_vs_filtered}
\end{figure}

{\color{black}{On short times compared to $\tau_{\text{s}}$ (or small length scales
compared to $\ell_{\text{s}}$) the link between $S$ and $w$ is weaker and the condensation reversibility is broken [see e.g.,~\citet{KC_1999a}]. This
irreversibility allows $w$ to generate supersaturation fluctuations
$S$ that may produce strictly local DSD broadening. These processes occurring on
time scales $\lesssim \tau_{\text{s}}$ and length scales $\lesssim \ell_{\text{s}}$ are resolved (although
coarsely represented) by the eddy hopping model if Eqs.~\eqref{eq:sde_aprime} to~\eqref{eq:dz_prime} are numerically integrated using a short time step $\delta t \ll \tau_{\text{s}}$. This is shown in the scatterplots of $z'$ and $a'$ in Fig.~\ref{fig:fine_vs_filtered}.a for finite $\tau_{s}$ and $\ell_{\text{s}}$, where a fine-grained DSD (red line) of finite breadth [measured by $\sigma_{a}(0)$] can be observed. The fine-grained DSD in the steady state is however much narrower than the filtered DSD (gray line) for a filter width $\Delta = 50 \: \text{m}$, as shown in Fig.~\ref{fig:fine_vs_filtered}.b.}}   

\begin{figure}
    \centering
\includegraphics[width=1.0\linewidth]{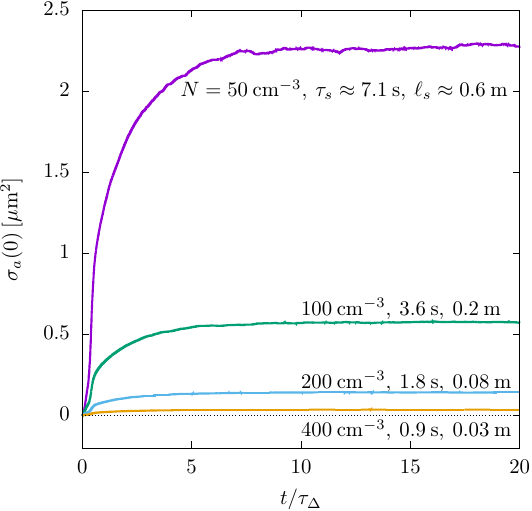}
    \caption{Fine-grained $\sigma_{a}(0)$ for different droplet number concentrations $N$ (the corresponding $\tau_{s}$ and $\ell_{s}$ in the steady state are also indicated). Larger $\tau_{s}$ (and $\ell_{s}$) leads to broader fine-grained DSD. Simulation parameters are the same as in Fig.~\ref{fig:fine_vs_filtered}, except for~$N$.}
    \label{fig:fine_dsd_vs_conc}
\end{figure}

Also,
we may argue that inside a (generally small) volume of order $\sim
\ell^{3}_{s}$, droplet sizes and droplet vertical positions are
approximately uncorrelated. This feature has been noticed
by~\citet{Clark_and_Hall_1979} in the region of the activation layer,
where the relaxation time $\tau_{s}$ and its associated length scale
$\ell_{s}$ are enlarged compared to the conditions in the cloud
core. As one advances towards the cloud core, inhomogeneities on
scales larger than $\ell_{s}$ dominate and the regime of
(large-scale) nearly reversible condensation is restored. {\color{black}{Figure~\ref{fig:fine_dsd_vs_conc} shows the increase of the breadth $\sigma_{a}(0)$ of the fine-grained DSD with $\tau_{s}$ (and $\ell_{s}$). The time constant $\tau_{s}$ and the corresponding length $\ell_{s}$ were changed by considering different droplet number concentrations $N$ as indicated in Fig.~\ref{fig:fine_dsd_vs_conc}.}} Thus in a scenario where supersaturation fluctuations $S$ (and the
resulting variability in droplet growth conditions) arises from turbulent vertical velocity fluctuations only, effective local broadening of the fine-grained DSD can emerge.

\begin{figure}
    \centering
    \includegraphics[width=1.0\linewidth]{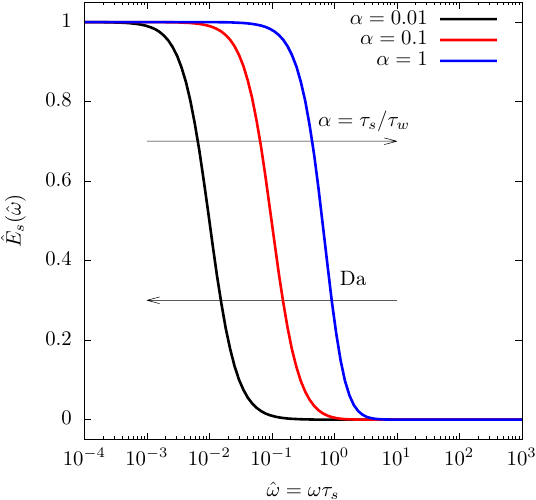}
    \caption{Dimensionless power spectrum $\hat{E}_{s}(\hat{\omega})$ of supersaturation fluctuations as a function of the dimensionless frequency $\hat{\omega} = \omega \tau_{s}$ for different values of the parameter $\alpha = \tau_{s}/\tau_{w}$. The arrow to the left indicates the direction of increasing Damk\"ohler number $\text{Da} \equiv \tau_{\text{mix}}/\tau_{\text{ph}}$.}
\label{fig:s_spectrum}
\end{figure}

{\color{black}{
Finally, we note that an alternative picture can be provided by examining the power spectrum $E_{s}(\omega)$ of the supersaturation fluctuation $S(t)$ driven by vertical velocity fluctuations $w(t)$. In Appendix~B we derive the expression for $E_{s}(\omega)$ predicted by the eddy hopping model. 

Figure~\ref{fig:s_spectrum} shows the dimensionless spectrum $\hat{E}_{s}(\hat{\omega})$ [Eq.~\eqref{eq:s_spectrum_dimensionless}] as a function of the dimensionless frequency $\hat{\omega} = \omega \tau_{s}$ for different values of $\alpha = \tau_{s}/\tau_{w}$, the ratio of the integral time scales of $S(t)$ and $w(t)$. As $\tau_{s} = 1/(\tau^{-1}_{\text{ph}} + \tau^{-1}_{\text{mix}}) \simeq \min \{\tau_{\text{ph}},\tau_{\text{mix}} \}$ and $\tau_{\text{mix}} \approx \tau_{w}$, the parameter $\alpha$ is closely related to the inverse of the Damk\"ohler number $\text{Da}$, defined as $\text{Da} \equiv \tau_{\text{mix}}/\tau_{\text{ph}}$. It is clear from Fig.~\ref{fig:s_spectrum} that the major contribution to the variance $\sigma^{2}_{s} = \int_{0}^{\infty} E_{s}(\omega)d\omega$, and hence to the variance $\sigma^{2}_{a}$ of droplet squared radius, comes from the lower frequency range. As the ratio $\alpha = \tau_{s}/\tau_{w}$ increases ($\text{Da}$ decreases), this frequency range shifts towards larger frequencies.

The high-frequency cutoff $\omega_{\text{c}}$ (defined so that
$\int_{0}^{\omega_{\text{c}}}E_{s}(\omega) d\omega$ approximates
$\sigma^{2}_{s}$ well enough) is specified by the time step $\delta t$
used to integrate the stochastic eddy-hopping model
equations~\eqref{eq:sde_aprime}-\eqref{eq:dz_prime} numerically. This
time step should satisfy the condition $\delta t \ll \tau_{s}$. In the
so-called DNS-like models of type A (to be discussed in
Sec.~\ref{sec:dns}), $\omega_{\text{c}}$ is specified by the spatial
resolution $\delta x$ of the computational grid. Accordingly, the
spatial resolution should be fine enough to satisfy the condition
$\delta x \ll
\ell_{s}$~\citep{Clark_and_Hall_1979}. Figure~\ref{fig:s_spectrum}
suggests that the high-frequency cutoff $\omega_{\text{c}}$ increases
as $\alpha$ increases ($\text{Da}$ decreases).  }}

\section{DNS-like studies}
\label{sec:dns}

Direct numerical simulations (DNS) studies\footnote{Here and
henceforth DNS actually means ``DNS-like" studies as they also include
the so-called scaled-up DNS of~\citep{Thomas_et_al_2020} and implicit
large eddy simulations (ILES) studies
of~\citet{Grabowski_Thomas_acp_2021} and
~\citet{Grabowski_et_al_DNS_CCN_single,Grabowski_et_al_DNS_CCN_distr}.}
of turbulent condensation may be separated into two types based on how
supersaturation fluctuations are created:
\begin{itemize}
\item[(A)] by local vertical velocity fluctuations via the adiabatic cooling effect, or
\item[(B)] by random forcing of the thermodynamic scalars (temperature
  $T$ and vapor mixing ratio $q$, or supersaturation $S$) applied
  isotropically on the large scales of the computational box of size
  $\Delta$.
\end{itemize}
Appendix~C summarizes the equations comprising the DNS models of type
A and~B. They resolve the velocity and thermodynamic scalar fields
more faithfully than the simple stochastic models described in
Sec.~\ref{sec:parcel}. However, the eddy-hopping and the reduced
models serve as useful prototypes of DNS models of type~A and~B,
respectively.

\subsection{DNS-like models of type B}

DNS studies of type~B neglect the adiabatic cooling effect and simulate supersaturation fluctuations solely produced by the scalar
forcing at large scales $\sim \Delta$. This
scalar forcing mimics the effects of external boundary
conditions that sustain scalar fluctuations against turbulent mixing. The works
by~\citet{Paoli_JAS_2009},~\citet{siewert_bec_krstulovic_2017}~and~\citet{Saito_et_al_2019}
belong to this type of DNS studies, which actually model turbulent condensation in an unbounded and homogeneous cloud. {\color{black}{General features of DNS models of type~B can be understood by analyzing the reduced stochastic model discussed in Sec.~\ref{sec:parcel}.\ref{sec:reduced_stochastic_model}, provided $\sigma^{2}_{s}$ is due to sources other than vertical velocity fluctuations.}}

DNS models of type~B produce polydisperse droplet dispersions with
spatially homogeneous droplet size statistics as does the reduced
stochastic model of Sec.~\ref{sec:parcel}.\ref{sec:reduced_stochastic_model}. In this
scenario, usual periodic boundary conditions (PBC) in all directions are
applicable to mimic an unbounded system, where cloud boundaries are
imagined to be removed to infinity. The fine-grained DSDs coincide with filtered DSDs (for arbitrary filter width) and these models are thus free from artifacts that could mask the real origin of the observed local DSD broadening. This is not the case of DNS model of type~A, as we shall discuss below. 

\subsection{DNS-like models of type~A}

DNS studies of type~A do not apply any large-scale scalar forcing, but rely on the local adiabatic cooling effect (driven by vertical velocity
fluctuations) as the only source of supersaturation variability. Their purpose is to confine the problem of turbulent condensation to a study of the effect of ``local" or ``internal" sources of thermodynamic
variability caused by velocity fluctuations on a range of length scales smaller or of the order of $\Delta$. The works
by~\citet{Celani_et_al_2005,Celani_et_al_2008,Celani_et_al_2009,Lanotte_et_al_2009,Sardina_et_al_2015,Sardina_et_al_2018,Saito_and_Gotoh_2018,Thomas_et_al_2020,Grabowski_Thomas_acp_2021,Grabowski_et_al_DNS_CCN_single,Grabowski_et_al_DNS_CCN_distr} belong
to this rather problematic class of DNS models.

DNS models of type~A give rise to polydisperse droplet dispersions with droplet
size statistics that is spatially inhomogeneous in the vertical
direction as in the stochastic eddy hopping model. In this case, the periodic computational box containing the droplets should be regarded as a small sample embedded in a cloud which has a mean squared radius gradient in the vertical direction (Sec.~\ref{sec:parcel}.\ref{sec:spurious_broadening}). Despite being
ill-suited for this scenario, usual PBC (Fig.~\ref{fig:usual_vs_modified_pbc}.a) in the vertical direction have been
the rule in DNS models of type~A, thus causing numerical artifacts.

Issues with DNS studies of this type were generally suggested
by~\citet{Prabhakaran_et_al_2022}. Their argument was formulated in
terms of ``inadequate representation of vertical turbulent transport"
and on the fact that ``all the cloud droplets have the same life time"
in these idealized frameworks. Here we refine the statements
of~\citet{Prabhakaran_et_al_2022} and explain in more detail the sources of the numerical artifacts in DNS models of type~A.

\subsection{Artifacts in DNS models of type~A}

The problem with DNS studies of type~A arises from misuse of PBC in
the direction (vertical) along which the statistics of droplet sizes
is spatially inhomogeneous. Usual PBC (illustrated in Fig.~\ref{fig:usual_vs_modified_pbc}.a) takes a cloud droplet that had left
the computational box, say, through the top, and puts it back at the
bottom of the domain (keeping the droplet size). If these droplets
trajectories (corrected by usual PBC) were ``unfolded", then it
would reveal that droplets actually diffuse in the vertical due to
turbulent fluctuations. Thus the actual averaging length scale
$\tilde{\Delta}$ grows beyond the computational box size $\Delta$
(where momentum forcing is applied) and follows the scaling $\sim
t^{1/2}$ as in the eddy hopping model operating in adiabatic
parcels (see Sec.~\ref{sec:parcel} and illustrated in
Fig.~\ref{fig:parcel_boundary_conditions}.a.\footnote{Note that models
producing local fluctuations in $T$ (or $S$) by fluctuations in $w$
via the adiabatic cooling effect implicitly assume that air parcels
ascending in the computational box expand due to the vertical decrease
of the base-state hydrostatic pressure. {\color{black}{This vertical gradient in the base-state pressure evidently brakes the translational invariance within the DNS computational box}}. This
inhomogeinity is absent in DNS studies of type~B.}

\begin{figure}
    \centering
    \includegraphics[width=1.0\linewidth]{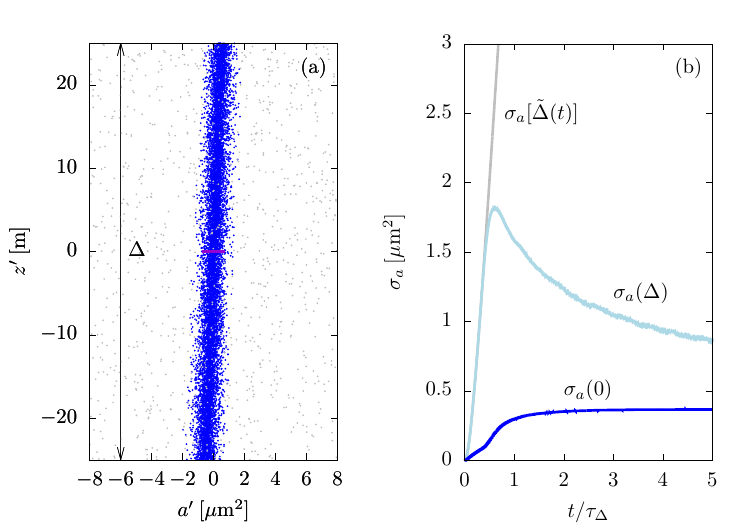}
    \caption{(a) Scatterplot of $a'$ and $z'$ using usual PBC (gray dots) and modified PBC (blue dots) for a background average supersaturation $\av{S} = 0.01$. Usual PBC artificially decorrelates $a'$ and $z'$ by mixing small and large droplets in the computational box. (b) Time evolution of the standard deviations $\sigma_{a}[\tilde{\Delta}(t)]$ (usual PBC), $\sigma_{a}(\Delta)$ and $\sigma_{a}(0)$ (corresponding to the filtered and fine-grained DSDs obtained under modified PBC). The results were obtained by numerical integration of the eddy-hopping model equations (a prototype for DNS-like models of type A). The scale height [Eq.~\eqref{eq:scale_height_a}] is $H_{a} \approx 7 \: \text{m}$ at $t/\tau_{\Delta} = 5$ and reaches $H_{a} \approx 18 \: \text{m}$ in the quasi-steady state.}
    \label{fig:usual_pbc_in_dns_typeA}
\end{figure}

Alternatively, the issue can be understood in the reversible
condensation scenario, which emerges on the large scales ($\gg
\ell_{s}$) inside the DNS computational box. We argue that the usual 
PBC, which returns a large droplet (that had left the computational
box at the top) back to the bottom of the domain (and keeps its large
radius), artificially brakes the large-scale near reversibility of the
condensation. It enforces that large and small droplets coexist in the computational box of characteristic size $\Delta$ and incorrectly
predicts DSD continuous broadening, which cannot be regarded as ``local". {\color{black}{This is shown in Fig.~\ref{fig:usual_pbc_in_dns_typeA}. The gray dots in Fig.~\ref{fig:usual_pbc_in_dns_typeA}.a show that usual PBC decorrelates $a$ and $z$ by vertically mixing small and large droplets in the computational box. The corresponding breadth $\sigma_{a}[\tilde{\Delta}(t)]$ grows continuously in time (Fig.~\ref{fig:usual_pbc_in_dns_typeA}.b). Conversely,  modified PBC preserves the vertical gradient of the droplet squares radius (blue dots in Fig.~\ref{fig:usual_pbc_in_dns_typeA}.a). The breadths $\sigma_{a}(\Delta)$ and $\sigma_{a}(0)$ of the corresponding filtered and fine-grained DSDs reach quasi-stationary values with $\sigma_{a}(\Delta) > \sigma_{a}(0)$ (Fig.~\ref{fig:usual_pbc_in_dns_typeA}). 

Also, usual PBC in DNS models of type A vertically mixes large and small droplets and creates spurious spatial correlations between droplets of different sized at short separations. This may artificially enhance droplet collision probabilities, which is essential for the onset of collision-coalescence. We shall discuss this further in Sec.~\ref{sec:sgs_model}.\ref{sec:coalescence_sgs}. For instance, the observation in DNS data by~\citep{Celani_et_al_2005} that small and large droplets cohabit in small volumes (and that it may enhance droplet collision probabilities) might have been distorted by the artificial effects of usual PBC. 

Finally, we note that besides using modified PBC we have to use sufficiently high grid resolutions to properly predict fine-grained DSDs and the spatial correlations between ``large" and ``small" droplets. The analysis
of~\citet{Clark_and_Hall_1979} and our discussion in
Sec.~\ref{sec:clark_hall} suggest that actual local DSD broadening can be observed only on small length scales compared to the supersaturation relaxation length scale $\ell_{s}$. Then the DNS grid needs to resolve small scales $\lesssim \ell_{s}$, as anticipated
by~\citet{Clark_and_Hall_1979}. The local broadening arising
from the irreversible coupling between supersaturation fluctuations
and the vertical motion of air parcels is expected to be enhanced around the activation layer (larger $\tau_{s}$ and $\ell_{s}$) and suppressed under cloud core conditions (smaller $\tau_{s}$ and $\ell_{s}$) as indicated in Fig.~\ref{fig:fine_dsd_vs_conc}. }}

\subsection{Sedimentation and droplet inertia}

Another important question is the treatment of droplet transport by
the turbulent flow. If we assume the droplets follow the turbulent air
flow $\mbf u(\mbf x,t)$ --- the droplet-tracer limit ---, thus the
droplet trajectory $\mbf X$ is defined via $\partial_{t} \mbf X = \mbf
u[\mbf X(t),t]$. Otherwise, if we cannot neglect the effects of
droplet inertia (characterized by the Stokes number $\text{St}$) and
sedimentation (characterized by the settling number $\text{Sv}$), then
$\mbf X(t)$ is determined via
\begin{equation}
    \partial_{t} \mbf X = \mbf u[\mbf X(t),t] + \text{corrections}(\text{St},\text{Sv}),
\label{eq:droplet_transport}
\end{equation}
where the corrections to the droplet-tracer limit depend on the Stokes
and settling numbers.

In DNS studies of type~A that additionally assume the droplet-tracer
limit (as the eddy hopping model), the microphysics is maximally
coupled with the droplet transport (the positive correlation between
droplets sizes and their vertical positions is maximized). As written,
ill-suited usual PBC in the vertical direction artificially and thus
unphysically brakes this correlation. Droplet inertia and
sedimentation work to weaken the correlation between microphysics and
droplet transport in the scenario of $w$-driven microphysical
variability. However, DNS studies that solve an equation
like~(\ref{eq:droplet_transport}) for the droplet transport
[e.g.,~\citet{Lanotte_et_al_2009,Celani_et_al_2005}] have considered
mainly regimes of small $\text{St}$ and $\text{Sv}$, where the the
so-called crossing-trajectory effects arising from sedimentation and
droplet inertia are negligible.

\section{Eddy hopping as a subgrid scheme in LES}
\label{sec:sgs_model}

If used as an SGS condensation scheme in LES with particle-based
microphysics, the eddy hopping model will predict filtered DSD that
tend to be broader than the those prognosed under the uniform
condensation approximation, where all droplets in a particular grid
box experience the same filtered (grid-averaged)
supersaturation~$\av{S}$. This has some implications that we shall
discuss here.

First, to prevent the eddy hopping model from predicting artificial
DSD broadening (through the non-local sampling discussed for the
adiabatic cloud parcel), the SGS turbulent transport of cloud droplets
must be treated properly. Second, if the SGS scheme predicts a broader
filtered DSD, then this will certainly have an impact on the filtered
condensation rate, and hence on the filtered buoyancy
production. Finally, broader filtered DSDs may (at least in principle)
indicate enhanced probabilities of collision-coalescence of droplets
due to differential sedimentation. However, enhancement of collision
probabilities actually occurs only if the physics underlying the SGS
scheme is able to produce strictly local DSD broadening, which
manifests itself mainly in the fine-grained DSD and not only in the
filtered DSD.

\subsection{SGS droplet transport}

If the eddy hoping model is used as a subgrid scheme in LES, then it
must provide microphysical properties that refer to a particular grid
box of size $\Delta$, and hence, these properties must be sampled over
a local averaging volume of characteristic size $\Delta$. (Here we
assume for simplicity that the computational grid is isotropic with
the same spatial resolution $\Delta$ in all directions.) To enforce
this \textit{local sampling} in LES, the SGS vertical transport of
droplets must be driven by the same SGS velocity fluctuations $w$
driving SGS supersaturation fluctuations $S$.

Otherwise, if the SGS model for condensation (including fluctuations
in $S$ driven by $w$) is decoupled from the SGS droplet transport,
then it will spuriously overestimate the local width of the filtered
DSD. For example, this is the risk we run while implementing in LES an
SGS scheme like the simplified stochastic model suggested
by~\citet{Saito_et_al_2021} (see
Sec.~\ref{sec:parcel}.\ref{sec:incomplete_equivalence}), where the SGS
vertical velocity fluctuation $w$ no longer appears as a resolved
model variable (although its underlying effects are parametrized in
the simplified model).

Even when the droplet SGS transport is properly treated, the local
variance of the filtered DSD may depend on the grid resolution
$\Delta$ (the filter width) if it is larger or of the same order of
$H_{a}$.

In an LES setting, $w$ is the vertical component of the unresolved
fluid velocity $\mbf u$, $w = \hat{\mbf e}_{z} \cdot \mbf u$. The
velocity fluctuation $\mbf u$ can be modeled by an stochastic differential equation of the form [see e.g.,~\citet{Pope_ARFM}]
\begin{equation}
d \mbf u = \bm \alpha \: dt - \frac{\mbf u}{\tau_{u}} \: d t + \sqrt{\frac{2
    \sigma^{2}_{u}}{\tau_{u}}} \: d \mbf W(t),
\label{eq:sde_uprime}
\end{equation}
where $d \mbf W(t)$ is the increment of an isotropic vector-valued
Wiener process. Equation~(\ref{eq:sde_uprime}) applies for the general
case of inhomogeneous turbulent flow in cloud scale simulations. It
differs from~(\ref{eq:dw_prime}), which is for idealized adiabatic
parcels filled with homogeneous turbulence, by the drift term [see e.g.,~\citet{SDE_velocity_fluctuations}],
\begin{equation}
\bm \alpha = \nabla \cdot \av{\mbf u \mbf u} - \mbf u \cdot \nabla \av{\mbf U},
\label{eq:drift}
\end{equation}
where $\av{\mbf U}$ is the filtered velocity field resolved by the
dynamical core of the cloud LES scheme, and $\av{\mbf u \mbf u}$ is
the kinematic Reynolds stress tensor. (Here, the spatial filtering
operation denoted by $\av{\cdot}$ is the same defined
in~(\ref{eq:filtered_dsd_def}) for the DSD and consider filter width
equal to the LES resolution length $\Delta$.)  Also, $\tau_{w}$ in
Eq.~(\ref{eq:dw_prime}) for stationary homogeneous turbulence is an
integral timescale, while $\tau_{u}$ {\color{black}{in~\eqref{eq:sde_uprime}}} for
inhomogeneous turbulence represents a local velocity decorrelation
timescale [see e.g.,~\citet{Rodean_book,Pope_simple_models}]. In the
droplet-tracer regime the droplet position $\mbf X$ is determined by
integrating the SDE $d\mbf X = [\av{\mbf U} + \mbf u] d t$.
Equations~(\ref{eq:sde_uprime}) and~(\ref{eq:drift}) constitute the
simplified Langevin model (SLM) for the fluid particle velocity
fluctuation used in several applications {\color{black}{[see
  e.g.,~\citet{Pope_ARFM,SDE_velocity_fluctuations}]}}.

A scheme that couples $w$ driving supersaturation fluctuations and
vertical SGS droplet transport was recently implemented in cloud scale
simulations by \citet{Chandrakar_et_al_2021_SGS}. They use an
expression derived by~\citet{Weil_et_al_2004} for the drift term $\bm
\alpha$, which is an approximation to
Eq.~(\ref{eq:drift}). Although~(\ref{eq:drift}) is essentially
``kinematic" (i.e., it arises as a kinematic step while writing the
equation of motion for the fluctuation $\mbf u$ from the NS equations
for the full velocity $\mbf U = \av{\mbf U} + \mbf u $), the
approximated drift term used by~\citet{Weil_et_al_2004}
and~\citet{Chandrakar_et_al_2021_SGS} makes this feature less
transparent. Also,~\citet{Chandrakar_et_al_2021_SGS} included an
additional random forcing to the stochastic equation for the SGS
Lagrangian supersaturation fluctuation, parametrized by a prescribed
supersaturation standard deviation
$\sigma_{s}$. Although~\citet{Chandrakar_et_al_2021_SGS} argue that
this additional forcing arises from sources other than vertical
velocity fluctuations, they specify $\sigma_{s}$ in terms of the
velocity fluctuation parameters~$\sigma_{u}$ and $\tau_{u}$. {\color{black}{A more physically appealing route to parametrize this stochastic forcing term of
supersaturation fluctuations has been considered in~\citet{Chandrakar_etal_jas_2023} as we discussed in Sec.~\ref{sec:parcel}.\ref{sec:reduced_stochastic_model}.}}

Finally, we briefly remark on the Lagrangian cloud model L3
by~\citet{Hoffman_2019a} and~\citet{Hoffman_et_al_2019b}. In this
model, the SGS droplet transport is not linked to the SGS vertical
velocity fluctuations that partially contribute to the subgrid
microphysical variability (through differential adiabatic cooling
rates experienced by the cloud droplets). The L3 model combines an LES
dynamical core with a particle-based microphysics, where the so-called
linear eddy model (LEM) [see
  e.g.,~\citet{Kerstein_LEM_1988,Krueger_et_al_1997}] is used to
represent SGS turbulent mixing of the thermodynamic scalars. LEM
solves the molecular diffusion of scalars on a one-dimensional grid
(the LEM grid), which is periodically rearranged and compressed (using
the so-called triple map) to simulate turbulent advection. As stated
by~\citet{Hoffman_2019a} and ~\citet{Hoffman_et_al_2019b}, LEM also
accounts for the SGS microphysical variability arising from
variability of adiabatic cooling rates experienced by different
droplets in a given LES grid box. This arises from stochastic vertical
motions due to the LEM triplet map, because the droplets experience
the same rearrangements as the boxes of the LEM grid, which is aligned
vertically. However, the actual SGS vertical droplet transport (which
is due to stochastic velocities computed from the SGS turbulent
kinetic energy $k$) and the stochastic vertical motions of the LEM
grid boxes (due to the LEM triplet map) are apparently uncorrelated.

\subsection{Grid-averaged condensation rate}

In an LES setting of spatial resolution $\Delta$, the balance equation
for the filtered vapor mixing ratio $\av{q}$ has the form
\begin{equation}
\partial_{t} \av{q} = - \av{C} + \cdots, 
\label{eq:filtered_q}
\end{equation}
where $\av{C}$ is the spatially filtered (or grid-averaged)
condensation rate. (Again, the spatial filtering $\av{\cdot}$ is the
same as in~(\ref{eq:filtered_dsd_def}) for the DSD.) The dots on the
RHS of~(\ref{eq:filtered_q}) denote other contributions to local
changes in $\av{q}$, such as advection by the turbulent flow and
molecular diffusion of water vapor.

In particle-based schemes [see e.g.,~\citet{Shima_2009,Arabas_et_al_2015,Dziekan_et_al_UWLCM_2019,Hoffman_2019a}], the filtered
condensation rate is calculated directly. If we neglect for
simplicity surface tension and solute effects and assume each droplet
grows according to~(\ref{eq:growth_eq_a}), then $\av{C}$ may be
approximated as
\begin{equation}
\av{C} \approx \alpha \sum_{k} r_{k} S_{k}, \; \; \; \; \alpha =
\frac{4 \pi \rho_{\text w} D}{\rho_{\text d} V_{\Delta}},
\end{equation}
where the sum is over all cloud droplets in the grid box of volume
$V_{\Delta}$, $r_{k}$ is the radius of the $k$th droplet experiencing
a supersaturation $S_{k}$; $\rho_{\text w}$ and $\rho_{\text d}$ are
the densities of water and dry air, respectively. Also, we assume that
the effective vapor diffusion coefficient $D$ varies weakly in
$V_{\Delta}$.

By decomposing
\begin{equation}
r_{k} = \av{r} + r'_{k}, \; \; \;  S_{k} = \av{S} + S'_{k},  
\end{equation}
where $\av{\cdot}$ denotes averages over all droplets in $V_{\Delta}$
at a particular time, we may write
\begin{equation}
\av{C} = \alpha N [ \av{r}\av{S} + \av{r'S'}],
\label{eq:condensation_rate}
\end{equation}
where $N$ is the total droplet concentration in the grid box. Thus the
condensation rate $\approx \alpha N \av{r}\av{S}$ under the uniform
condensation approximation is increased by a factor proportional to
the covariance $\av{r'S'}$ if this is positive. For collisionless and
non-sedimenting droplets, a positive correlation between $S$ and $r$
simply indicates that the large droplet tail of the filtered DSD is
formed by those droplets that experienced higher supersaturation along
their growth histories.

An equation like~(\ref{eq:condensation_rate}) was considered
by~\citet{Paoli_JAS_2009}. They stated that the \textit{closure
  problem} of determining the filtered condensation term $\av{C}$
boils down to {\color{black}{modeling the covariance $\av{r'S'}$ (i.e., writing it in terms of model resolved variables)}}. The eddy hopping
model used as a particle-based SGS scheme provides a ``closure" by
explicitly calculating $\av{C}$ beyond the uniform condensation
approximation. The resulting broadening of the filtered DSD predicted
by the eddy hopping model renders $\av{r'S'} > 0$ and leads to
increased local condensation rates compared with those obtained under
uniform condensation. Enhanced filtered condensation rates does not
require locally broader fine-grained DSDs; broader filtered DSDs
suffice. In contrast, enhanced droplet collision probabilities are
supported by broader fine-grained DSDs only, as we shall discuss next.

\subsection{Stochastic droplet coalescence}
\label{sec:coalescence_sgs}

Although we are primarily focused on the DSD formed by the growth of
collisionless droplets through condensation, the resulting DSD impacts
the onset of gravity-induced droplet collision-coalescence.

Particle-based schemes for cloud microphysics, such as the
superdroplet method (SDM) developed by~\citet{Shima_2009}, treat
droplet collision-coalescence in a probabilistic manner through a
Monte Carlo method. This requires the probability $P_{ij}$ that two
droplets (labeled $i$ and $j$) will coalesce  {\color{black}{inside a volume $V$ during the time interval $(t,t + \delta t)$.}} This probability is given by
\begin{equation}
P_{ij} = K_{ij} \delta t /V,
\label{eq:p_ij}
\end{equation}
where $K_{ij}$ is the coalescence kernel. For droplet collisions
induced by gravity through differential sedimentation,
\begin{equation}
K_{ij} \propto
(r_{i} + r_{j})^{2} |w_{\text{sed}}(r_{i})-w_{\text{sed}}(r_{j})|,
\label{eq:k_ij}
\end{equation}
where $w_{\text{sed}}(r)$ is a parametrization for the sedimentation
velocity of a droplet of radius $r$.

\begin{figure*}
    \centering
    \includegraphics[width=0.7\linewidth]{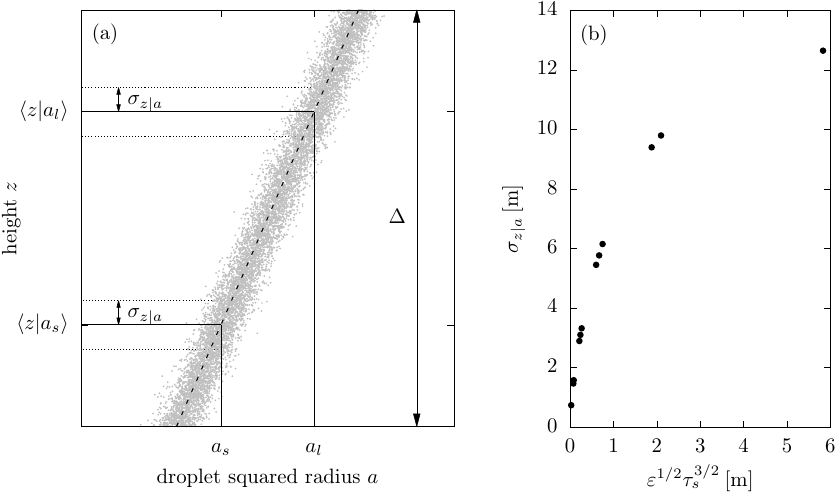}
    \caption{(a) Scatterplot of $a$ and $z$ showing the conditional expected heights $\av{z|a_{s}}$ and $\av{z|a_{l}}$, respectively, of a ``small" droplet (of squared radius $a_{s}$) and a ``large" droplet ($a_{l}$). Considering the indicated conditional dispersion $\sigma_{z|a}$ in heights, these ``large" and ``small" droplets are in fact well-separated (and \textit{not} thoroughly mixed) in the volume of characteristic length $\Delta$. (b) Conditional standard deviation $\sigma_{z|a}$ of heights (for a given $a$) as a function of the supersaturation length scale $\ell_{s} \sim \varepsilon^{1/2} \tau^{3/2}_{s}$. The points are for $\varepsilon$ in the range $0.001-0.1 \: \text{m}^{2}\text{s}^{-3}$ and droplet number concentration $N$ in the range $50-400 \: \text{cm}^{-3}$.}
    \label{fig:spatial_correlation}
\end{figure*}

An essential assumption underlying~(\ref{eq:k_ij}) is that the
colliding droplets are well-mixed inside $V$ [see e.g.,~\citet{Gillespie_1972,Shima_2009,Dziekan_Pawlowska_ACP_coalescence}]
as spatial correlations between droplets is not parametrized in the
collision kernel~\eqref{eq:k_ij}\footnote{\color{black}{Consider a ``small" droplet of radius, say, $r_{s} = \av{r}_{\Delta} - 2\sigma_{r}(\Delta)$ and a ``large" one of radius $r_{l} = \av{r}_{\Delta} + 2\sigma_{r}(\Delta)$ are located at the same grid box of size $\Delta$, where the average radius $\av{r}_{\Delta}$ and the standard deviation $\sigma_{r}(\Delta)$ refer to the filtered DSD $\av{n(r)}_{\Delta}$. Now assume $N_{l}$ is the number concentration of large droplets. We define the radial distribution function $g_{sl}$ for small and large droplets such that $N_{l} g_{sl}(\mbf x) d\mbf x$
is the (ensemble) average number of large droplets in a volume element $d\mbf x$ at the position $\mbf x$ from a small droplet. The gravity collision kernel~\eqref{eq:k_ij} assumes $g_{sl}(d \hat{\mbf x}) = 1$, where $d = r_{s} + r_{l}$ and $\hat{\mbf x} = \mbf x/|\mbf x|$, while in fact $g_{sl}(d\hat{\mbf x}) \approx 0$.}}. In an LES setting, it is generally assumed that the
volume $V$ in~(\ref{eq:p_ij}) is associated with the model spatial
resolution, hence with the volume $V_{\Delta}$ of the computational
grid box. Compared with the uniform condensation approximation, the
eddy hopping model used as an SGS condensation scheme tends to predict
relatively broad filtered DSD (of underlying sampling volume $\sim
V_{\Delta}$). However, large and small droplets forming the left and
right tails of the predicted filtered DSD {\color{black}{might not be}} well-mixed in the
volume $V_{\Delta}$ of the grid box. This is illustrated by scatterplots of $a$ and $z$ in Fig.~\ref{fig:spatial_correlation}. In fact, at the early stages of
condensation, large and small droplets may be well separated in the
computational grid box {\color{black}{(as sedimentation is still too slow to effectively mix large and small droplets).}} This may lead to conceptual difficulties in
calculating collision probabilities according to~(\ref{eq:k_ij}).

As discussed in Sec.~\ref{sec:clark_hall}, the eddy hopping model
(relying on $S$ driven solely by $w$) presumably produces droplet
size statistics that are spatially homogeneous only on scales smaller
or of the order of the supersaturation relaxation length scale
$\ell_{s}$. Then one may argue that the well-mixed volume $V$ has a
characteristic linear size $V^{1/3} \lesssim \ell_{s}$, which may be much
smaller (e.g., under cloud core conditions) than the volume $V_{\Delta}$ of
the computational grid boxes. Also, $\av{n(r)}_{\ell_{s}}$ (the
filtered DSD with filter width $\sim \ell_{s}$) is much narrower than
$\av{n(r)}_{\Delta}$ (the filtered DSD with filter width $\sim
\Delta$). Thus droplet collision probabilities in $V \sim
\ell_{s}^{3}$ may not be significantly enhanced (at least in the cloud
core) under the turbulent condensation represented by the eddy hopping
model. {\color{black}{Figure~\ref{fig:spatial_correlation}.b shows how the dispersion $\sigma_{z|a}$ around the expected height $\av{z|a}$ of a droplet of squared radius $a$ depends on $\ell_{s}$. As $\ell_{s}$ increases, the heights of a ``small" and a ``large" droplet are more broadly distributed, thus enhancing the probability of finding droplets of different sizes close to each other.}}  

\section{Conclusions}
\label{sec:conclusions}

In this paper we discussed that any turbulent condensation model that
(in some degree) relates supersaturation fluctuations $S$ experienced
by cloud droplets with their vertical velocity fluctuations $w$ will
produce correlations between droplet sizes and droplet vertical
positions in a cloudy volume. As a consequence, the statistics of
droplet sizes produced by such a model will be spatially inhomogeneous
along the vertical direction. This implies that the droplet size
distribution (DSD) will depend on the spatial scale over which the
distribution is sampled. To systematically discuss this dependence on
the sampling spatial scale, we have defined (Sec.~\ref{sec:dsd}) the
fine-grained DSD $n(r)$ and the filtered DSD $\av{n(r)}_{\Delta}$ of
filter width $\Delta$. Thus when the droplet size statistics is
spatially inhomogeneous, then fine-grained DSDs that are locally
narrow may
render filtered DSDs that are locally broad in a discretized domain of spatial resolution $\Delta$ if $\Delta$ is sufficiently large compared to $H_{a}$ [Eq.~\eqref{eq:scale_height_a}].

A simple model that links $S$ with $w$ is the stochastic eddy
hopping
model~\citep{Sardina_et_al_2015,Grabowski_Abade_2017,Abade_et_al_2018},
where vertical velocity fluctuations causes supersaturation variability through the adiabatic cooling
effect. The idea behind the eddy hopping model is that, in a
relatively large cloudy volume (say, of the size of a typical LES grid
box), various cloud droplets (that are representative of the whole
droplet population in this large cloudy volume) may experience
different adiabatic cooling rates depending on the vertical turbulent
velocity of the air carrying these droplets.

Used as a stand-alone representation of turbulent condensation in
adiabatic parcel framework, the original version of the eddy hopping
model produces broadening of the parcel DSD that continuously grow in
time. We have shown that this continuous growth does not reflect an
actual local DSD broadening [as supposed by~\citet{Sardina_et_al_2015,Grabowski_Abade_2017,Abade_et_al_2018}], but arises from the ``non-local" sampling caused by the growth with time (like $\sim t^{1/2}$) of the spatial scale (or filter width)
associated with the sampling volume containing the whole droplet
population of the parcel. The sampling volume is implicitly related to the spatial extent of the droplet turbulent diffusion in the vertical direction, which is driven by the same vertical velocity fluctuations that cause
variability in the supersaturation experienced by different droplets. {\color{black}{To enforce local sampling in the adiabatic parcel framework,
    we considered (Sec.~\ref{sec:parcel}.\ref{sec:spurious_broadening}) alternative boundary conditions (reflecting and modified PBC) at the top and bottom horizontal boundaries of the sampling volume (see Sec.~\ref{sec:spurious_broadening}).}}

A certain type of DNS-like models of
turbulent condensation
[e.g.,~\citet{Celani_et_al_2005,Celani_et_al_2008,Celani_et_al_2009,Lanotte_et_al_2009,Sardina_et_al_2015,Sardina_et_al_2018,Saito_and_Gotoh_2018,Thomas_et_al_2020,Grabowski_Thomas_acp_2021,Grabowski_et_al_DNS_CCN_single,Grabowski_et_al_DNS_CCN_distr}],
as they create supersaturation fluctuations through local vertical
velocity fluctuations, has the same artifacts as the much simpler
stochastic eddy hopping model. Hence the broad DSD these DNS-like
studies predict cannot be regarded as strictly local. Free of such
inconsistencies are other DNS studies
[e.g.,~\citet{Paoli_JAS_2009,siewert_bec_krstulovic_2017,Saito_et_al_2019}],
where supersaturation fluctuations are sustained by random scalar
forcing applied on the largest scales of the computational box
(supersaturation fluctuations at small scales are determined by
downscale nonlinear transfer). 

The eddy hopping model can be used as a subgrid condensation scheme in
LES (with particle-based microphysics) and tends to predict broad
filtered DSD relative to the uniform condensation approximation. We
emphasized in this work that actual local sampling of the filtered DSD must be enforced in LES by properly coupling the subgrid
droplet transport with the subgrid vertical velocity fluctuations that
cause subgrid supersaturation fluctuations.

Broader filtered DSD predicted by the eddy hopping model may increase
local (grid-averaged) condensation rates in LES compared with uniform
condensation. However, broad filtered DSD does not necessarily imply
enhanced probabilities of droplet collisions induced by differential
sedimentation (as broad filtered DSD does not imply broad fine-graned
DSD). Broader fine-grained DSD can be supported by the eddy hopping
model only through the condensation reversibility breaking that may
occur at small scales of the order of the supersaturation relaxation
lengthscale $\ell_{s}$ [see Sec.~\ref{sec:clark_hall} and the analysis
  of~\citet{Clark_and_Hall_1979}]. However, the filtered DSD
$\av{n(r)}_{\ell_{s}}$ with filter width $\sim \ell_{s}$ may be narrow in
the cloud core (under the turbulent condensation represented by the
eddy hopping model) and cannot effectively enhance droplet collision
probabilities.

The analysis presented in this paper neglects the effects of droplet
inertia and sedimentation. As these effects deviate droplets from
Lagrangian fluid trajectories, droplet inertia and sedimentation
affect the Lagrangian correlation time $\tau_{s}$ for the
supersaturation fluctuations and work to destroy correlations between
droplet sizes and droplet vertical positions. Clarifying these effects
poses several challenges to stochastic representations of turbulent
condensation, such as the eddy hopping model. However, this problem
can be tackled by DNS-like models, as done, for instance, by~\citet{Vaillancourt_et_al_2002}.

 It should be evident that the eddy hopping model (and the DNS-like
 models where $S$ is driven by $w$) cannot explain the stationary
 DSD obtained from simulation data of the Pi Chamber setup, as those
 reported and analyzed by~\citet{Prabhakaran_et_al_2022}. The
 variability of the droplet growth conditions in the Pi Chamber and in
 the idealized models considered here are attributed to different
 sources, operating on different spatial scales. In the Pi Chamber
 [see e.g.,~\citet{Pi_Chamber_BAMS}], the variability of the droplet growth conditions are caused by fixed scalar boundary
 conditions (for temperature and vapor on the chamber walls) that
 sustain supersaturation fluctuations against isobaric mixing due to
 Rayleigh-B\'enard turbulence. Also, the domain of the Pi Chamber is
 of relatively small vertical extent compared with typical grid boxes
 in LES. This rules out any significant variability in the adiabatic
 cooling rates among droplets (ascending with different turbulent
 vertical velocities), which is the main source of variability in the
 droplet growth conditions postulated by the models discussed in this
 work. Finally, we note that the turbulent condensation models
 discussed here assume the cloudy volume is effectively unbounded
 (although the DSD sampling volume is finite) and neglect droplet
 transport by sedimentation. In the Pi Chamber (which may be regarded
 as a homogeneous but bounded system) large droplets are removed by
 sedimentation from the sampling volume. This truncates the right tail
 of the sampled DSD, despite supersaturation fluctuations being
 continuously maintained by the boundary conditions.

\clearpage

\acknowledgments

I am grateful to Christoph Siewert
for his stimulating criticism of the eddy hopping model that motivated this paper. This manuscript benefited from discussions with Prasanth Prabhakaran
and Raymond Shaw and was initially drafted during my visit to the
Kavli Institute for Theoretical Physics (KITP) at the University of
California Santa Barbara, that is sponsored by the National Science
Foundation under Grant No. NSF PHY-1748958. I thank Wojciech Grabowski for discussions and for his comments on an earlier
version of this paper. I am indebted
to Marta Waclawczyk who drew my attention to the simplified Langevin
model discussed in Sec.~\ref{sec:sgs_model}. We thank two anonymous reviewers for their comments on the manuscript. This work was partly
supported by the Polish National Science Centre (NCN) under Grant
No. 2017/25/B/ST10/02383.

%
%
\datastatement


The data that support the findings of this study are
available from the author upon reasonable
request.

%





\appendix[A]
\appendixtitle{Modified periodic boundary conditions}
\label{sec:modified_pbc}

We sketch here a simple procedure to apply the modified PBC shown schematically in Fig.~\ref{fig:usual_vs_modified_pbc}.b. The scheme resembles the so-called Lees-Edwards boundary conditions~\citep{Lees_Edwards_1972} used in particle-based molecular dynamics simulations of fluids under shear flow. 

Knowing the droplet vertical position $z(t)$ and its squared radius $a(t)$ at time $t$, first obtain the provisional values $z_{\ast}$ and $a_{\ast}$ at time $t + \delta t$ by numerical integration of~\eqref{eq:dz_prime} and~\eqref{eq:da_dz_cooper} over the model time step $\delta t$. Then correct (if necessary) the provisional values for periodicity to obtain $z(t + \delta t)$ and $a(t + \delta t)$ as follows,
\begin{equation}
z(t + \delta t) = z_{\ast} + \epsilon \: \Delta, 
\end{equation}
and 
\begin{equation}
a(t + \delta t) = a_{\ast} + \epsilon \: \gamma_{a} \Delta , 
\label{eq:rescaling_a}
\end{equation}
where
\begin{equation}
\epsilon = \left \{
\begin{array}{ll}
+1, & \text{if} \; z_{\ast} < \av{z} - \Delta/2, \\
-1, & \text{if} \; z_{\ast} > \av{z} + \Delta/2, \\
0, & \text{otherwise.}
\end{array}
\right .
\end{equation}
is the periodicity correction index, $\av{z}$ is the mean vertical position of the droplets of the ensemble, and $\bar{\gamma}_{a}$ is the mean vertical gradient of the squared radius $a$.  

For the eddy hopping model operating in adiabatic parcels we may assume that the vertical structure of $a$ follows the quasi-steady profile [Eq.~\eqref{eq:da_propto_dz}] over the length scale $\Delta$ of the sampling volume and approximate
\begin{equation}
\bar{\gamma}_{a} \approx 2 D A_{1} \tau_{s}, 
\label{eq:gamma}
\end{equation}
where the phase relaxation time $\tau_{\text{ph}}$ in $\tau_{s}$ is approximated as
\begin{equation}
\tau_{\text{ph}} \approx 1/[A_{2}\av{\mu_{1}}_{\Delta}].
\label{eq:tau_bar}
\end{equation}
For DNS-like models of type~A, instead of using~\eqref{eq:gamma} and~\eqref{eq:tau_bar}, $\bar{\gamma}_{a}$ may be estimated from a least square fit to a linear profile of the droplet squared radius. 

\appendix[B]
\appendixtitle{Phase modification and supersaturation spectrum}
\label{sec:s_spectrum}

\begin{figure}[h!]
    \centering
    \includegraphics[width=1.0\linewidth]{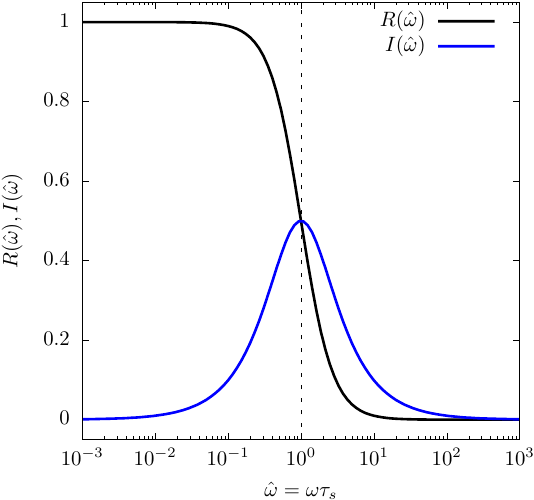}
    \caption{Reduced functions $R(\hat{\omega})$ and $I(\hat{\omega})$ as a function of the dimensionless frequency $\hat{\omega} = \omega \tau_{s}$.}
\label{fig:reduced_functions}
\end{figure}

The Fourier-Laplace transforms $S(\omega)$ and $w(\omega)$ of $S(t)$ and $w(t)$ are linearly related through the transfer function $\varphi(\omega) = \varphi'(\omega) + i \varphi''(\omega)$ [Eq.~\eqref{eq:s_frequency}].
It is of interest to write $\varphi'(\omega)$ and $\varphi''(\omega)$ in terms of reduced functions of the dimensionless frequency $\hat{\omega} = \omega \tau_{s}$, 
\[ R(\hat{\omega}) \equiv \frac{\varphi'(\tilde{\omega})}{\varphi(0)} = \frac{1}{1 + \hat{\omega}^{2}}, \; \; \; \; I(\hat{\omega}) \equiv \frac{\varphi''(\hat{\omega})}{\varphi(0)} = \frac{\hat{\omega}}{1 + \hat{\omega}^{2}}. \]
The reduced functions are plotted in Fig.~\ref{fig:reduced_functions}. The real part $R(\hat{\omega})$, which is in phase with the ``forcing" $w(t)$, dominates the ``response" $S(t)$ in the lower frequency range ($\hat{\omega} = \omega \tau_{s} \lesssim 1$), but vanishes in the high-frequency regime ($\hat{\omega} \gtrsim 10$). The phase modification due to $I(\hat{\omega})$ takes place on a range of frequencies around $1/\tau_{s}$ (that is, $\hat{\omega} = 1$). 

Now let us consider the power spectrum $E_{s}(\omega)$ of supersaturation fluctuations $S(t)$ driven by vertical velocity fluctuations $w(t)$. It can be shown that $E_{s}(\omega)$ is related to the power spectrum $E_{w}(\omega)$ of $w(t)$ through the transfer function $\varphi(\omega)$ defined in~\eqref{eq:s_frequency},
\begin{equation} 
E_{s}(\omega) = |\varphi(\omega)|^{2} E_{w}(\omega).
\label{eq:es_ew_relation}
\end{equation}
The statistically stationary process $w(t)$ is described by Eq.~\eqref{eq:dw_prime} and has autocorrelation function 
\[ R_{w}(t) = \av{w'(t_{0})w'(t_{0}+t)} = \sigma^{2}_{w} e^{-|t|/\tau_{w}},  \]
with integral time scale $\tau_{w}$. The spectrum $E_{w}(\omega)$ is twice the Fourier transform of $R_{w}(t)$ and equals [see e.g.,~\citet{Pope_book}]
\begin{equation} 
E_{w}(\omega) = \frac{1}{\pi} \int_{-\infty}^{\infty} R_{w}(t) e^{-i\omega t} dt = \frac{2}{\pi} \frac{\sigma^{2}_{w} \tau_{w}}{1 + \omega^{2} \tau^{2}_{w}}.
\label{eq:ew}
\end{equation}
Accordingly, $E_{s}(\omega)$ is twice the Fourier transform of $R_{s}(t) = \av{S'(t_{0})S'(t_{0}+t)}$ with integral time scale $\tau_{s}$. Then it follows from~\eqref{eq:es_ew_relation}, ~\eqref{eq:s_frequency}, and~\eqref{eq:ew} that 
\[ E_{s}(\omega) = \frac{2}{\pi} \frac{A^{2}_{1} \tau^{2}_{s}}{1 + \omega^{2} \tau^{2}_{s}} \frac{\sigma^{2}_{w} \tau_{w}}{1 + \omega^{2} \tau^{2}_{w}}. \]
The dimensionless spectrum $\hat{E}_{s}(\hat{\omega})$, scaled by $(2/\pi) A^{2}_{1}\tau^{2}_{s}\sigma^{2}_{w}\tau_{w}$, in terms of the dimensionless frequency $\hat{\omega} = \omega \tau_{s}$ is, 
\begin{equation}
\hat{E}_{s}(\hat{\omega}) = \frac{1}{1 + \hat{\omega}^{2}} \frac{1}{1+(\hat{\omega}/\alpha)^{2}},
\label{eq:s_spectrum_dimensionless}
\end{equation}
where $\alpha = \tau_{s}/\tau_{w}$ is the ratio of the integral time scales of $S(t)$ and $w(t)$. 

\appendix[C]
\appendixtitle{Equations for DNS-like models}
\label{sec:dns_model_equations}

In general, DNS studies of turbulent condensation solve the
Navier-Stokes (NS) equations\footnote{Some studies
[e.g.,~\citet{Paoli_JAS_2009}] solve the equations for compressible
fluid motion. However, this feature is not relevant for the present
discussion.} for the velocity field $\mbf u$,
\begin{equation}
\partial_{t} \mbf u = \cdots + \: \mbf f, \; \; \; \; \; \nabla \cdot \mbf u = 0,
\end{equation}
where on the right-hand side (RHS) we omit the convective acceleration
and the well-known NS terms, making explicit only the external
statistically homogeneous and isotropic momentum forcing $\mbf f$. It
is applied on the larger scales $\sim \Delta$ of the computational box
to maintain a turbulent stationary flow. This momentum forcing mimics
turbulence production by scales larger than $\Delta$ and is a common
feature of all DNS studies.

The treatment of thermodynamics and moisture, however, differ among
various DNS studies. Some models resolve both the temperature $T$ and
the vapor mixing ratio $q$ as the thermodynamic prognostic variables. The local
changes of $q$ and $T$ are described by equations of the form,
\begin{equation}
\partial_{t} q = f_{q} + \cdots, 
\label{eq:dns_q}
\end{equation}
and
\begin{equation}
\partial_{t} T = - \Gamma w + f_{T} + \cdots, 
\label{eq:dns_temp}
\end{equation}
where $\Gamma = g/c_{p}$ is the lapse rate accounting for the
adiabatic cooling effect, $w = \mbf u \cdot \hat{\mbf e}_{z}$ is the
local vertical velocity, $f_{q}$ and $f_{T}$ are large-scale
stochastic forcing of the vapor and temperature fields. The dots on
the RHS of~(\ref{eq:dns_q}) and~(\ref{eq:dns_temp}) represent other
contributions, such as advection by the resolved turbulent field $\mbf
u$, molecular diffusion and condensation (accompanied by latent heat
exchanges). The local supersaturation $S$ specifying the ambient
conditions for droplet growth is diagnosed from $T$ and $q$.

Other DNS studies do not resolve $T$ or $q$, but directly compute the
supersaturation $S$ driving droplet growth by condensation. In this
$S$-formulation for the thermodynamic scalars, the supersaturation
obeys an equation of the form
\begin{equation}
\partial_{t} S =  A_{1} w + f_{S} + \cdots, 
\label{eq:dns_s}
\end{equation}
where $f_{S}$ stands for the large-scale supersaturation forcing and $A_{1}$ is the coefficient [same as in Eq.~(\ref{eq:quasi-steady-approx})] accounting for the adiabatic cooling effect. Again, the dots on the RHS of~(\ref{eq:dns_s}) represent turbulent advection by $\mbf u$, molecular diffusion and condensation. The RHS of the equations above for $T$, $q$, and $S$ makes explicit only the ``external'' stochastic sources producing fluctuations in the thermodynamics scalars (other than small-scale fluctuations due to the nonlinear cascade) that lead to variability in droplets growth conditions.

DNS studies of type~B neglect the adiabatic cooling effect by assuming $\Gamma = 0$ (or $A_{1} = 0$ in the $S$-formulation), and simulate supersaturation fluctuations solely produced by the scalar
forcing at large scales $\sim \Delta$; hence they assume $f_{T} \not = 0$ and $f_{q} \not = 0$ in~(\ref{eq:dns_temp}) and~(\ref{eq:dns_q})
[or $f_{S} \not = 0$ in the $S$-formulation~(\ref{eq:dns_s})].

DNS studies of type~A do not apply any large-scale scalar forcing. Hence they assume $f_{T} = f_{q} = 0$ in~(\ref{eq:dns_temp})
and~(\ref{eq:dns_q}) [or $f_{S} = 0$ in~(\ref{eq:dns_s})] and rely on
the local adiabatic cooling effect driven by vertical velocity fluctuations ($\Gamma \not= 0$ or $A_{1} \not=0$) as the only source of supersaturation variability.

%



\bibliographystyle{ametsocV6}
\bibliography{references}

\end{document}